\def\today{\ifcase\month\or January\or
February\or March\or April\or May\or June\or
  July\or August\or September\or October\or November\or December\fi
  \space\number\day, \number\year}
\documentclass{emulateapj}
\newcommand{\chimin}{2.936}
\newcommand{\imin}{60.6}
\newcommand{\kmin}{0.087}
\newcommand{\PEQmin}{8.784}	
\newcommand{\UOThreemin}{1.0011}
\newcommand{\EOThreeONEmin}{1410.93}
\newcommand{\pOThreeONEmin}{9.008}
\newcommand{\latOThreeONEmin}{32.4}
\newcommand{\gOThreeONEmin}{11.75}
\newcommand{\EOThreeTWOmin}{1406.10}
\newcommand{\pOThreeTWOmin}{9.070}
\newcommand{\latOThreeTWOmin}{37.2}
\newcommand{\gOThreeTWOmin}{5.95}
\newcommand{\UOFourmin}{1.0149}
\newcommand{\EOFourONEmin}{1756.68}
\newcommand{\pOFourONEmin}{8.820}
\newcommand{\latOFourONEmin}{12.7}
\newcommand{\gOFourONEmin}{7.99}
\newcommand{\EOFourTWOmin}{1763.54}
\newcommand{\pOFourTWOmin}{9.200}
\newcommand{\latOFourTWOmin}{-46.3}
\newcommand{\gOFourTWOmin}{16.76}
\newcommand{\EOFourTHREEmin}{1769.09}
\newcommand{\pOFourTHREEmin}{9.572}
\newcommand{\latOFourTHREEmin}{77.3}
\newcommand{\gOFourTHREEmin}{13.04}
\newcommand{\UOFivemin}{1.0042}
\newcommand{\EOFiveONEmin}{2116.94}
\newcommand{\pOFiveONEmin}{9.372}
\newcommand{\latOFiveONEmin}{58.4}
\newcommand{\gOFiveONEmin}{9.74}
\newcommand{\EOFiveTWOmin}{2120.22}
\newcommand{\pOFiveTWOmin}{9.248}
\newcommand{\latOFiveTWOmin}{49.6}
\newcommand{\gOFiveTWOmin}{8.55}
\newcommand{\vsinimin}{4.77}
\newcommand{\eqspeedmin}{5.47}

\newcommand{\chiLow}{2.968} 
\newcommand{\chiHigh}{3.018} 
\newcommand{\iLow}{57.8} 
\newcommand{\iHigh}{63.5} 
\newcommand{\kLow}{0.085} 
\newcommand{\kHigh}{0.096} 
\newcommand{\kBest}{0.090} 
\newcommand{\kPlus}{0.006} 
\newcommand{\kMinus}{0.005} 
\newcommand{\PEQLow}{8.74} 
\newcommand{\PEQHigh}{8.81} 
\newcommand{\PEQBest}{8.77} 
\newcommand{\PEQPlus}{0.03} 
\newcommand{\PEQMinus}{0.04} 
\newcommand{\vsiniLow}{4.64} 
\newcommand{\vsiniHigh}{4.94} 
\newcommand{\eqspeedLow}{5.46} 
\newcommand{\eqspeedHigh}{5.50} 
\newcommand{\UOThreeLow}{1.0003} 
\newcommand{\UOThreeHigh}{1.0017} 
\newcommand{\EOThreeONELow}{1410.92} 
\newcommand{\EOThreeONEHigh}{1410.94} 
\newcommand{\pOThreeONELow}{8.990} 
\newcommand{\pOThreeONEHigh}{9.015} 
\newcommand{\latOThreeONELow}{29.5} 
\newcommand{\latOThreeONEHigh}{34.8} 
\newcommand{\gOThreeONELow}{11.63} 
\newcommand{\gOThreeONEHigh}{11.86} 
\newcommand{\EOThreeTWOLow}{1406.07} 
\newcommand{\EOThreeTWOHigh}{1406.16} 
\newcommand{\pOThreeTWOLow}{9.022} 
\newcommand{\pOThreeTWOHigh}{9.094} 
\newcommand{\latOThreeTWOLow}{32.9} 
\newcommand{\latOThreeTWOHigh}{39.8} 
\newcommand{\gOThreeTWOLow}{5.68} 
\newcommand{\gOThreeTWOHigh}{6.18} 
\newcommand{\UOFourLow}{1.0129} 
\newcommand{\UOFourHigh}{1.0150} 
\newcommand{\EOFourONELow}{1756.66} 
\newcommand{\EOFourONEHigh}{1756.70} 
\newcommand{\pOFourONELow}{8.787} 
\newcommand{\pOFourONEHigh}{8.851} 
\newcommand{\latOFourONELow}{9.3} 
\newcommand{\latOFourONEHigh}{16.8} 
\newcommand{\gOFourONELow}{7.73} 
\newcommand{\gOFourONEHigh}{8.09} 
\newcommand{\EOFourTWOLow}{1763.49} 
\newcommand{\EOFourTWOHigh}{1763.56} 
\newcommand{\pOFourTWOLow}{9.153} 
\newcommand{\pOFourTWOHigh}{9.231} 
\newcommand{\latOFourTWOLow}{-47.6} 
\newcommand{\latOFourTWOHigh}{-43.2} 
\newcommand{\gOFourTWOLow}{14.44} 
\newcommand{\gOFourTWOHigh}{17.31} 
\newcommand{\EOFourTHREELow}{1769.03} 
\newcommand{\EOFourTHREEHigh}{1769.16} 
\newcommand{\pOFourTHREELow}{9.542} 
\newcommand{\pOFourTHREEHigh}{9.640} 
\newcommand{\latOFourTHREELow}{74.9} 
\newcommand{\latOFourTHREEHigh}{78.4} 
\newcommand{\gOFourTHREELow}{11.60} 
\newcommand{\gOFourTHREEHigh}{13.53} 
\newcommand{\UOFiveLow}{1.0029} 
\newcommand{\UOFiveHigh}{1.0051} 
\newcommand{\EOFiveONELow}{2116.95} 
\newcommand{\EOFiveONEHigh}{2117.05} 
\newcommand{\pOFiveONELow}{9.365} 
\newcommand{\pOFiveONEHigh}{9.423} 
\newcommand{\latOFiveONELow}{55.4} 
\newcommand{\latOFiveONEHigh}{62.1} 
\newcommand{\gOFiveONELow}{9.24} 
\newcommand{\gOFiveONEHigh}{10.28} 
\newcommand{\EOFiveTWOLow}{2120.21} 
\newcommand{\EOFiveTWOHigh}{2120.30} 
\newcommand{\pOFiveTWOLow}{9.166} 
\newcommand{\pOFiveTWOHigh}{9.252} 
\newcommand{\latOFiveTWOLow}{42.9} 
\newcommand{\latOFiveTWOHigh}{50.1} 
\newcommand{\gOFiveTWOLow}{7.94} 
\newcommand{\gOFiveTWOHigh}{8.52} 

\newcommand{\chiMeLow}{3.145} 
\newcommand{\chiMeHigh}{3.374} 
\newcommand{\iMeLow}{44.9} 
\newcommand{\iMeHigh}{74.1} 
\newcommand{\kMeLow}{0.094} 
\newcommand{\kMeHigh}{0.117} 
\newcommand{\PEQMeLow}{8.71} 
\newcommand{\PEQMeHigh}{8.90} 
\newcommand{\vsiniMeLow}{3.83} 
\newcommand{\vsiniMeHigh}{5.34} 
\newcommand{\eqspeedMeLow}{5.40} 
\newcommand{\eqspeedMeHigh}{5.53} 
\newcommand{\UOThreeMeLow}{0.9964} 
\newcommand{\UOThreeMeHigh}{1.0031} 
\newcommand{\EOThreeONEMeLow}{1410.94} 
\newcommand{\EOThreeONEMeHigh}{1410.97} 
\newcommand{\pOThreeONEMeLow}{8.978} 
\newcommand{\pOThreeONEMeHigh}{9.007} 
\newcommand{\latOThreeONEMeLow}{16.7} 
\newcommand{\latOThreeONEMeHigh}{36.9} 
\newcommand{\gOThreeONEMeLow}{11.05} 
\newcommand{\gOThreeONEMeHigh}{11.93} 
\newcommand{\EOThreeTWOMeLow}{1405.85} 
\newcommand{\EOThreeTWOMeHigh}{1406.23} 
\newcommand{\pOThreeTWOMeLow}{8.948} 
\newcommand{\pOThreeTWOMeHigh}{9.160} 
\newcommand{\latOThreeTWOMeLow}{-45.7} 
\newcommand{\latOThreeTWOMeHigh}{58.9} 
\newcommand{\gOThreeTWOMeLow}{5.08} 
\newcommand{\gOThreeTWOMeHigh}{9.41} 
\newcommand{\UOFourMeLow}{1.0025} 
\newcommand{\UOFourMeHigh}{1.0092} 
\newcommand{\EOFourONEMeLow}{1756.69} 
\newcommand{\EOFourONEMeHigh}{1756.78} 
\newcommand{\pOFourONEMeLow}{8.721} 
\newcommand{\pOFourONEMeHigh}{8.939} 
\newcommand{\latOFourONEMeLow}{-0.8} 
\newcommand{\latOFourONEMeHigh}{15.0} 
\newcommand{\gOFourONEMeLow}{7.17} 
\newcommand{\gOFourONEMeHigh}{8.51} 
\newcommand{\EOFourTWOMeLow}{1763.53} 
\newcommand{\EOFourTWOMeHigh}{1763.65} 
\newcommand{\pOFourTWOMeLow}{9.039} 
\newcommand{\pOFourTWOMeHigh}{9.316} 
\newcommand{\latOFourTWOMeLow}{-43.9} 
\newcommand{\latOFourTWOMeHigh}{-28.7} 
\newcommand{\gOFourTWOMeLow}{7.14} 
\newcommand{\gOFourTWOMeHigh}{19.21} 
\newcommand{\EOFourTHREEMeLow}{1769.01} 
\newcommand{\EOFourTHREEMeHigh}{1769.32} 
\newcommand{\pOFourTHREEMeLow}{9.572} 
\newcommand{\pOFourTHREEMeHigh}{9.751} 
\newcommand{\latOFourTHREEMeLow}{59.2} 
\newcommand{\latOFourTHREEMeHigh}{76.9} 
\newcommand{\gOFourTHREEMeLow}{7.53} 
\newcommand{\gOFourTHREEMeHigh}{13.23} 
\newcommand{\UOFiveMeLow}{1.0000} 
\newcommand{\UOFiveMeHigh}{1.0053} 
\newcommand{\EOFiveONEMeLow}{2116.96} 
\newcommand{\EOFiveONEMeHigh}{2117.09} 
\newcommand{\pOFiveONEMeLow}{9.340} 
\newcommand{\pOFiveONEMeHigh}{9.419} 
\newcommand{\latOFiveONEMeLow}{41.8} 
\newcommand{\latOFiveONEMeHigh}{63.0} 
\newcommand{\gOFiveONEMeLow}{8.30} 
\newcommand{\gOFiveONEMeHigh}{9.37} 
\newcommand{\EOFiveTWOMeLow}{2120.09} 
\newcommand{\EOFiveTWOMeHigh}{2120.21} 
\newcommand{\pOFiveTWOMeLow}{9.094} 
\newcommand{\pOFiveTWOMeHigh}{9.253} 
\newcommand{\latOFiveTWOMeLow}{28.0} 
\newcommand{\latOFiveTWOMeHigh}{56.0} 
\newcommand{\gOFiveTWOMeLow}{7.69} 
\newcommand{\gOFiveTWOMeHigh}{8.31} 

\newcommand{\RVALTwoSix}{1.012}
\newcommand{\RVALTwoFive}{1.053}
\newcommand{\RVALTwoFour}{1.010}
\newcommand{\RVALTwoZero}{1.072}
\newcommand{\RVALOneNine}{1.059}
\newcommand{\RVALOneTwo}{1.042}
\newcommand{\RVALOneOne}{1.098}
\newcommand{\RVALOneZero}{1.013}
\newcommand{\RVALThree}{1.026}
\newcommand{\RVALTwo}{1.023}
\newcommand{\RVALZero}{1.031}
\newcommand{\RVALOne}{1.026}
\newcommand{\RVALFour}{1.006}
\newcommand{\RVALFive}{1.009}
\newcommand{\RVALSix}{1.024}
\newcommand{\RVALSeven}{1.022}
\newcommand{\RVALEight}{1.100}
\newcommand{\RVALNine}{1.035}
\newcommand{\RVALOneThree}{1.004}
\newcommand{\RVALOneFour}{1.005}
\newcommand{\RVALOneFive}{1.039}
\newcommand{\RVALOneSix}{1.124}
\newcommand{\RVALOneSeven}{1.070}
\newcommand{\RVALOneEight}{1.092}
\newcommand{\RVALTwoOne}{1.057}
\newcommand{\RVALTwoTwo}{1.038}
\newcommand{\RVALTwoThree}{1.179}

\newcommand{\RVALParallelTwoSix}{1.477}
\newcommand{\RVALParallelTwoFive}{1.443}
\newcommand{\RVALParallelTwoFour}{1.033}
\newcommand{\RVALParallelTwoZero}{1.074}
\newcommand{\RVALParallelOneNine}{1.249}
\newcommand{\RVALParallelOneTwo}{1.490}
\newcommand{\RVALParallelOneOne}{1.596}
\newcommand{\RVALParallelOneZero}{1.338}
\newcommand{\RVALParallelThree}{1.548}
\newcommand{\RVALParallelTwo}{1.144}
\newcommand{\RVALParallelZero}{1.110}
\newcommand{\RVALParallelOne}{1.312}
\newcommand{\RVALParallelFour}{1.338}
\newcommand{\RVALParallelFive}{1.051}
\newcommand{\RVALParallelSix}{1.415}
\newcommand{\RVALParallelSeven}{1.126}
\newcommand{\RVALParallelEight}{1.209}
\newcommand{\RVALParallelNine}{1.237}
\newcommand{\RVALParallelOneThree}{1.179}
\newcommand{\RVALParallelOneFour}{1.391}
\newcommand{\RVALParallelOneFive}{1.079}
\newcommand{\RVALParallelOneSix}{1.318}
\newcommand{\RVALParallelOneSeven}{1.122}
\newcommand{\RVALParallelOneEight}{1.057}
\newcommand{\RVALParallelTwoOne}{1.182}
\newcommand{\RVALParallelTwoTwo}{1.289}
\newcommand{\RVALParallelTwoThree}{1.235}

\newcommand{\chiMeFourLow}{1.50} 
\newcommand{\chiMeFourHigh}{1.82}
\newcommand{\kMeFourLow}{0.095} 
\newcommand{\kMeFourHigh}{0.111}
\newcommand{\PEQMeFourLow}{8.624} 
\newcommand{\PEQMeFourHigh}{8.721}
\newcommand{\vsiniMeFourLow}{4.80} 
\newcommand{\vsiniMeFourHigh}{4.86}
\newcommand{\UOFourMeFourLow}{1.0051} 
\newcommand{\UOFourMeFourHigh}{1.0150} 
\newcommand{\EOFourONEMeFourLow}{1756.66} 
\newcommand{\EOFourONEMeFourHigh}{1756.74} 
\newcommand{\pOFourONEMeFourLow}{8.704} 
\newcommand{\pOFourONEMeFourHigh}{8.773} 
\newcommand{\latOFourONEMeFourLow}{9.1} 
\newcommand{\latOFourONEMeFourHigh}{22.3} 
\newcommand{\gOFourONEMeFourLow}{7.51} 
\newcommand{\gOFourONEMeFourHigh}{7.80} 
\newcommand{\EOFourTWOMeFourLow}{1763.40} 
\newcommand{\EOFourTWOMeFourHigh}{1763.50} 
\newcommand{\pOFourTWOMeFourLow}{9.050} 
\newcommand{\pOFourTWOMeFourHigh}{9.162} 
\newcommand{\latOFourTWOMeFourLow}{-45.8} 
\newcommand{\latOFourTWOMeFourHigh}{-38.6} 
\newcommand{\gOFourTWOMeFourLow}{11.21} 
\newcommand{\gOFourTWOMeFourHigh}{15.80} 
\newcommand{\EOFourTHREEMeFourLow}{1768.94} 
\newcommand{\EOFourTHREEMeFourHigh}{1769.07} 
\newcommand{\pOFourTHREEMeFourLow}{9.518} 
\newcommand{\pOFourTHREEMeFourHigh}{9.598} 
\newcommand{\latOFourTHREEMeFourLow}{71.0} 
\newcommand{\latOFourTHREEMeFourHigh}{77.9} 
\newcommand{\gOFourTHREEMeFourLow}{9.04} 
\newcommand{\gOFourTHREEMeFourHigh}{13.11} 

\newcommand{\kFourBest}{0.096} 
\newcommand{\PEQFourBest}{8.68} 
\newcommand{\chiFourLow}{1.46} 
\newcommand{\chiFourHigh}{1.52}
\newcommand{\kFourLow}{0.092} 
\newcommand{\kFourHigh}{0.101}
\newcommand{\PEQFourLow}{8.65} 
\newcommand{\PEQFourHigh}{8.72}
\newcommand{\vsiniFourLow}{4.80} 
\newcommand{\vsiniFourHigh}{4.84}
\newcommand{\UOFourFourLow}{1.0139} 
\newcommand{\UOFourFourHigh}{1.0150} 
\newcommand{\EOFourONEFourLow}{1756.64} 
\newcommand{\EOFourONEFourHigh}{1756.68} 
\newcommand{\pOFourONEFourLow}{8.736} 
\newcommand{\pOFourONEFourHigh}{8.788} 
\newcommand{\latOFourONEFourLow}{15.6} 
\newcommand{\latOFourONEFourHigh}{21.7} 
\newcommand{\gOFourONEFourLow}{7.64} 
\newcommand{\gOFourONEFourHigh}{7.84} 
\newcommand{\EOFourTWOFourLow}{1763.43} 
\newcommand{\EOFourTWOFourHigh}{1763.49} 
\newcommand{\pOFourTWOFourLow}{9.112} 
\newcommand{\pOFourTWOFourHigh}{9.183} 
\newcommand{\latOFourTWOFourLow}{-47.5} 
\newcommand{\latOFourTWOFourHigh}{-43.9} 
\newcommand{\gOFourTWOFourLow}{14.65} 
\newcommand{\gOFourTWOFourHigh}{16.83} 
\newcommand{\EOFourTHREEFourLow}{1768.97} 
\newcommand{\EOFourTHREEFourHigh}{1769.08} 
\newcommand{\pOFourTHREEFourLow}{9.510} 
\newcommand{\pOFourTHREEFourHigh}{9.579} 
\newcommand{\latOFourTHREEFourLow}{76.6} 
\newcommand{\latOFourTHREEFourHigh}{78.0} 
\newcommand{\gOFourTHREEFourLow}{12.64} 
\newcommand{\gOFourTHREEFourHigh}{13.17} 

\newcommand{\RVALFourNine}{1.003}
\newcommand{\RVALFourEight}{1.002}
\newcommand{\RVALFourSeven}{1.000}
\newcommand{\RVALFourSix}{1.001}
\newcommand{\RVALFourFive}{1.003}
\newcommand{\RVALFourFour}{1.013}
\newcommand{\RVALFourThree}{1.003}

\newcommand{\RVALFourOne}{1.016}
\newcommand{\RVALFourZero}{1.002}
\newcommand{\RVALFourOneZero}{1.002}
\newcommand{\RVALFourOneOne}{1.000}
\newcommand{\RVALFourOneTwo}{1.011}

\newcommand{\RVALFourParallelNine}{1.127}
\newcommand{\RVALFourParallelEight}{1.034}
\newcommand{\RVALFourParallelSeven}{1.315}
\newcommand{\RVALFourParallelSix}{1.110}
\newcommand{\RVALFourParallelFive}{1.203}
\newcommand{\RVALFourParallelFour}{1.162}
\newcommand{\RVALFourParallelThree}{1.183}

\newcommand{\RVALFourParallelOne}{1.074}
\newcommand{\RVALFourParallelZero}{1.092}
\newcommand{\RVALFourParallelOneZero}{1.279}
\newcommand{\RVALFourParallelOneOne}{1.102}
\newcommand{\RVALFourParallelOneTwo}{1.122}

\slugcomment{Draft \today}

\shorttitle{Differential rotation of $\kappa^1$ Ceti}
\shortauthors{Walker et al.}

\begin{document}


\title{THE DIFFERENTIAL ROTATION OF $\kappa^1$ CETI AS OBSERVED BY {\it MOST}\footnotemark[1] ~-- II
}

\footnotetext[1]{Based on data from the MOST satellite, a Canadian Space Agency mission, jointly operated by Dynacon Inc., the University of Toronto Institute of
Aerospace Studies and the University of British Columbia with the assistance of the University of Vienna.} 

\author{ Gordon A.H. Walker\altaffilmark{2}, Bryce Croll\altaffilmark{3}, Rainer Kuschnig\altaffilmark{3}, Andrew Walker\altaffilmark{4}, Slavek M. Rucinski\altaffilmark{5}, Jaymie M. Matthews\altaffilmark{3},  David B. Guenther\altaffilmark{6}, Anthony F.J. Moffat\altaffilmark{7}, Dimitar Sasselov\altaffilmark{8}, Werner W. Weiss\altaffilmark{9}} 

\altaffiltext{2}{1234 Hewlett Place, Victoria, BC V8S 4P7, Canada;
gordonwa@uvic.ca}

\altaffiltext{3}{Dept. Physics \& Astronomy, UBC, 
6224 Agricultural Road, Vancouver, BC V6T 1Z1, Canada; croll@phas.ubc.ca, kuschnig@phas.ubc.ca, matthews@phas.ubc.ca}

\altaffiltext{4}{Sumus Technology Ltd.; arwalker@sumusltd.com}

\altaffiltext{5}{Dept. Astronomy \& Astrophysics, David Dunlap Obs., Univ. Toronto 
P.O.~Box 360, Richmond Hill, ON L4C 4Y6, Canada;
rucinski@astro.utoronto.ca}

\altaffiltext{6}{Department of Astronomy and Physics, St. Mary's University
Halifax, NS B3H 3C3, Canada;
guenther@ap.stmarys.ca}

\altaffiltext{7}{ D\'ept de physique, Univ de Montr\'eal 
C.P.\ 6128, Succ.\ Centre-Ville, Montr\'eal, QC H3C 3J7, and Obs du mont M\'egantic, Canada; moffat@astro.umontreal.ca}

\altaffiltext{8}{Harvard-Smithsonian Center for Astrophysics, 
60 Garden Street, Cambridge, MA 02138, USA;
sasselov@cfa.harvard.edu}

\altaffiltext{9}{Institut f\"ur Astronomie, Universit\"at Wien 
T\"urkenschanzstrasse 17, A--1180 Wien, Austria;
weiss@astro.univie.ac.at}

\begin{abstract}
We first reported evidence for differential rotation of $\kappa^1$~Ceti
in Paper I. In this paper we demonstrate that the differential rotation pattern closely matches that for the Sun.
This result is based on additional {\it MOST} (Microvariability \& Oscillations of STars) observations in 2004 and 2005,
to complement the 2003 observations discussed in Paper I. Using StarSpotz, a program developed specifically to analyze {\it MOST} photometry, we have solved for $k$, the differential rotation coefficient, and $P_{EQ}$, the equatorial
rotation period using the light curves from all three years. The spots
range in latitude from 10$^{\circ}$ to 75$^{\circ}$ and $k$ =
$\kBest^{+\kPlus}_{-\kMinus}$ -- less than the solar value but
consistent with the younger age of the star. $k$ is also well
constrained by the independent spectroscopic estimate of $v$sin$i$.
We demonstrate independently that the pattern of differential rotation with latitude in fact conforms to solar. 

Details are given of the parallel tempering formalism used in finding
the most robust solution which gives $P_{EQ}$ = $\PEQBest^{+\PEQPlus}_{-\PEQMinus}$ \ days -- smaller than that usually adopted, implying an age $<$ 750 My.  Our values of
$P_{EQ}$ and $k$ can explain the range of  
rotation periods determined by others by spots or activity at a variety of latitudes.
Historically, Ca II activity seems to occur consistently between
latitudes 50$^{\circ}$ and 60$^{\circ}$ which might indicate a
permanent magnetic feature.  Knowledge of $k$ and $P_{EQ}$ are key to
understanding the dynamo mechanism and rotation structure in the
convective zone as well assessing age for solar-type stars. We
recently published values of $k$ and $P_{EQ}$ for $\epsilon$ Eri based
on {\it MOST} photometry and expect to analyse {\it MOST} light curves
for several more spotted, solar-type stars.
 
\end{abstract}



\section{INTRODUCTION}

\label{intro}

Differential rotation and convection provide the engine for the solar dynamo (eg. \citet{Oss03}). The mechanism suggested by \citet{Leb41} 
for differential rotation seems to be the most widely accepted. Angular momentum is continually redistributed in the deep convective envelope which occupies some 30\% of the outer solar radius  resulting in nonuniform rotation because the convective turbulence is affected by the Coriolis force \citep{Kit05}. There is good evidence that a younger Sun would have rotated more rapidly and gradually lost angular momentum through magnetic coupling to the solar wind and there is currently a considerable theoretical effort to develope quantitative models of differential rotation in solar-type stars (see, for example, reviews by \citet{Kit05} and \citet{Tho03}).

For stars other than the Sun, the simultaneous detection of two or more
spots at different latitudes allows a direct measurement of differential
rotation. The {\it MOST} photometric satellite \citep{MOST}
offers continuous photometry of target stars with an unprecedented
precision for weeks at a time making it it possible to 
track spots accurately. 

We \citep{Ruc04} have already reported the detection of a pair of spots in the 2003 {\it MOST} light curve of $\kappa^{1}$ Ceti (HD~20630, HIP~15457, HR~996; $V=4.83$, $B-V=0.68$, G5V). The photometry was part of the satellite
commissioning and prior to a considerable improvement pointing in early 2004.
The larger spot had a well-defined 8.9 d rotation period while the smaller one
had a period $\geq$9.3 d. Only the latitude of
the larger spot could be determined with any confidence. Spectroscopic
observations by \citet{Shk03}, mostly from 2002, of the Ca~II K-line
emission reversals were also synchronized to a period of about 9.3 d.
\citet{Ruc04} also determined a stellar radii of $R_{*}$ = 0.95 $\pm$ 0.10 $R_{\bigodot}$.

To better decipher the spot activity and rotation of ${\kappa}^1$ Ceti
and to model spot distributions on other {\it MOST} targets, one of us, BC,
developed the program Starspotz (\citet{Cro06b}; \citet{Cro06a}). It was applied first to a 2005 {\it MOST} light curve covering three
rotations of $\epsilon$ Eri where two spots were detected at
different latitudes rotating with different periods.  From this, we
derived a differential rotation coefficient for $\epsilon$ Eri (see section 4),
$k=0.11^{+.03}_{-.02}$, in agreement with models by \citet{Bro04}
for young Sun-like stars rotating roughly twice as fast as the Sun.

In this paper, we apply Starspotz to the three separate {\it MOST} light curves of $\kappa^{1}$ Ceti -- from 2003, 2004 and 2005. We specifically solve for the astrophysically important elements of inclination, differential rotation, equatorial period and rotation speed and find that the differential rotation profile is identical to solar.We use parallel tempering methods to fully explore the realistic parameter space and identify the global $\chi^{2}$ minimum to the data-set that signifies a best-fitting unique starspot configuration. Markov Chain Monte Carlo (MCMC) methods are then used to explore in detail the parameter space around this global minimum, to define best-fitting values and their uncertainties and to explore possible correlations among the fitted and derived parameters.

\setcounter{footnote}{4}      % otherwise corrupted numbers

\section{ {\it MOST} PHOTOMETRY AND LIGHT CURVES}
\label{most}

The {\it MOST} satellite, launched on June 2003, is fully described
by \citet{MOST}. A 15/17.3 cm Rumak-Maksutov telescope
feeds two CCDs, one for tracking and the other for science, through a single, custom, broadband filter (350 -- 700 nm). Starlight from primary science targets ($V \leq 6$) 
is projected onto the science CCD as a fixed (Fabry) image of the telescope entrance pupil covering some 1500 pixels. The experiment was designed to detect photometric variations with periods of minutes at micro-magnitude precision and does not use comparison stars or flat-fielding for calibration. There is no direct connection to any photometric system. Tracking jitter was dramatically reduced early in 2004 to $\sim$1 arcsec which, subsequently, led to significantly higher photometric precision.

The observations reported here were reduced by RK. Outlying data points generated by poor tracking or cosmic ray hits were removed. At certain orbital phases, {\it MOST} suffers from parasitic light, mostly Earthshine, the amount and phase depending on the stellar coordinates, spacecraft roll and season of the year. Data are recorded for seven Fabry images adjacent to the target to track the background. The background signals were combined in a mean and subtracted from the target photometry. This procedure also corrected for bias, dark and background signals and their variations. Reductions basically followed the scheme already outlined by \citet{Ruc04}. 

\begin{deluxetable*}{ccccccc}
\tabletypesize{\footnotesize}
\tablecaption{{\it MOST} Observations of $\kappa^1$ Ceti
\label{TableObs}
}
\tablewidth{0pt}
\tablehead{
\colhead{year}	&\colhead{dates}	 &\colhead{total} &\colhead{duty cycle}	&\colhead{exp }&\colhead{cycle}&\colhead{$\sigma$\tablenotemark{a}}\\
\colhead{start}	&\colhead{HJD -- 2451545}	  &\colhead{days} &\colhead{\%} &\colhead{sec}&\colhead{sec}&\colhead{mmag}}
\startdata
% 2003 ~5 Nov& (2452)948.461 to 978.901    &   30.44  &  96 & 40   &     50   &     0.5~~\\
% 2004 15 Oct& (2453)293.578 to 313.989    &   20.41  &  99 & 30    &   35   &     0.30\\
% 2005 14 Oct& (2453)657.721 to 670.065    &   12.34 &   83   &     30    &    35     &   0.19\\
2003 ~5 Nov& 1403.461 to 1433.901    &   30.44  &  96 & 40   &     50   &     0.5~~\\
2004 15 Oct& 1748.578 to 1768.989    &   20.41  &  99 & 30    &   35   &     0.30\\
2005 14 Oct& 2112.721 to 2125.065    &   12.34 &   83   &     30    &    35     &   0.19\\
\enddata
\tablenotetext{a}{for data binned at the {\it MOST} orbital period of 101.413 min.}
\end{deluxetable*}

A log of the {\it MOST} observations of $\kappa^1$~Ceti for each of 2003, 2004 and 2005 is given in Table 1. For the light curve analysis in this paper, data were binned at the {\it MOST} orbital period of 101.413 min. Figure 1 displays full light curves for each of the three years with time given as JD-2451545 (heliocentric).  The solid line in the lower panel displays the difference between the two  most divergent background readings from the seven background Fabry images. There is no obvious correlation between structure in the differential background signal and the light curve. The complete light curve can be downloaded from the MOST Public Archive at www.astro.ubc.ca/MOST.

%\placefigure{fig1}
The formal rms point-to-point precison, $\sigma$, is given in Table 1.
The shapes of the light curves ultimately used for the StarSpotz
analysis depend to a degree on just how the background is removed.
Despite having the summed signal values from 7 adjacent Fabry images we
cannot interpolate a `true' background but subtraction of the mean
background is a consistent option. In each of the three data sets used
here the 7 background values are closely identical when averaged over
an orbit but the differences between the extreme backgrounds can be
significant when the ambient background is high (eg. pointing close to
the Moon) which is why that difference is shown in the lower panel for
each light curve.

We have assumed that the true errors in the individual orbital means
making up each light curve are proportional to the difference between
the maximum and minimum background values. These errors have been
arbitrarily scaled by one-fourth for the formal StarSpotz analysis and
these `errors' are shown as bars in Figures  4, 5, and 6.  The
individual bars correspond approximately to the standard deviation of
the background.

\section{StarSpotz}
\label{SecStarSpotz}

StarSpotz \citep{Cro06b} is a program based on SPOTMODEL
(\citealt{Rib03}; \citealt{Rib02}), that is designed to provide
a graphical user interface for photometric spot modeling. The
functionality included in StarSpotz is designed to take advantage
of the nearly continuous photometry returned by {\it MOST} to quantify
the non-uniqueness problem often associated with photometric spot
modeling.  Basically, StarSpotz uses the analytic models of
\citet{Budding} or \citet{Dorren} to model the drop in intensity caused by $N_{SPOT}$ circular, non-overlapping spots. The model
proposed in \citet{Budding} is used in this application.

Fitting can be performed by a variety of methods including a
Marquardt-Levenberg non-linear least squares algorithm (\citealt{Marq};
\citealt{Leven}; \citealt{Press}), or Markov Chain Monte Carlo (MCMC)
functionality (\citealt{Gregorybook}; \citealt{Gamerman};
\citealt{Gilks}).  The parameters of interest are the stellar
inclination angle, $i$, the unspotted intensity of the star, $U$, and
the period, epoch, latitude and angular size of the $a^{th}$ spot:
$p_{a}$, $E_{a}$, $\beta_{a}$, and $\gamma_{a}$.  \citet{Cro06b} have demonstrated how the Marquardt-Levenberg non-linear least squares algorithm could be used to fit the cycle to cycle variation in $\epsilon$ Eri.
\citet{Cro06a} investigated the same observations of $\epsilon$ Eri demonstrating that Markov Chain Monte Carlo functionality is
very helpful in defining correlations and possible degeneracies between
the fitted parameters.

The search for additional local minima in $\chi^{2}$ space indicating 
other plausible spot configurations that produce a reasonable fit to the
light-intensity curve is an important part of the modeling
because it allows one to quantify the uniqueness of a solution.  We argue that this task of completely searching parameter space for
other physically plausible local minima is better suited to a process
called parallel tempering (\citealt{Gregorypaper};
\citealt{Gregorybook}), discussed here in $\S$\ref{SecTemper}, rather than the
Uniqueness test methods in \citet{Cro06b}.

The present and subsequent releases of the freely available program StarSpotz, 
including the full source-code, executable, and documentation are available at:

www.astro.ubc.ca/MOST/StarSpotz.html

Release 3.0 includes the functionality discussed which allows one to fit multiple epochs of data as well as the parallel
tempering functionality. 

\section{The Spots on $\kappa^1$  Ceti}

As the {\it MOST} observations of $\kappa^1$  Ceti come from three
well-seperated epochs with independent starspot groups from epoch to epoch, the fitting process was considerably
more complicated than those detailed in \citet{Cro06b} and
\citet{Cro06a}. Rather than fitting for the period, $p$, of the
individual spots in each epoch we derive a common equatorial period,
$P_{EQ}$, and the differential rotation coefficient, $k$, among the three
epochs where these are given by:
\begin{equation} P_{\beta} = P_{{\rm EQ}}/(1-k~{\rm sin}^{2}\beta)
\end{equation} where $P_{\beta}$ is the period at latitude $\beta$.
There are five fewer parameters to be fitted because the
periods of the individual spots are no longer independent variables,
but completely defined  by $k$, $P_{EQ}$, and $\beta$. Analyses of solar differential rotation usually include terms in sin$^{4}\beta$ \citep{Kit05} but we did not expect to detect spots at such high latitudes where this term would be important. 

Common limb-darkening coefficients, $u$ = 0.684, and flux ratios
between the spot and unspotted photosphere, $\kappa_{\omega}$ = 0.22
were assumed. The former was based on the compilations of
\citet{Dia95}, while the latter comes from the canonical value for
sunspots ($T_{spot}\sim4000$ K, $T_{phot}\sim5800$ K). The 27
parameters to be fitted are summarized in Table \ref{TablePrior}.

Initial investigations of the light curves using the StarSpotz standard
spot model indicated that different numbers of spots were 
required for the various epochs.  The 2003 and 2005 {\it MOST}
data-sets could be well explained by two starspots rotating with
different periods, at different latitudes. The behaviour observed in
the 2004 {\it MOST} light curve, on the other hand, required a
third spot. Thus the following analysis
explicitly assumes two, three and two starspots for the 2003-2005 {\it
MOST} epochs, respectively. 

In this present application we use the same paradigm that was applied
to $\kappa^{1}$ Ceti by \citet{Ruc04}, and $\epsilon$ Eri by
\citet{Cro06a} and \citet{Cro06b} where the spot parameters ($E$,
$\beta$, $\gamma$) were assumed to be constant throughout the
observations, and that the cycle-to-cycle variability observed is due
solely to spots rotating differentially.  Other interpretations are
certainly plausible including circular spots with time-varying spot sizes
and latitudes such as those applied to IM Peg in \citet{Rib03}.  The
Maximum Entropy \& Tikhonov reconstruction technique of
\citet{Lanza98}, which assumes pixellated spots of various contrasts, is another
example. We feel that with the limitations of photometric spot
modeling, our assumption of unchanging spot parameters is the most
suitable for the timescale of the {\it MOST} light curves. It provides an
opportunity to determine a unique solution, as well as
recovering astrophysically important parameters.

We recognise, given the obvious changes in the $\kappa^{1}$ Ceti light
curve over a year, that the assumption of spot parameters constancy
over a few weeks is a simplification but we do not expect significant
spot migration on timescales of less than a month.

\subsection{StarSpotz Markov Chain Monte Carlo (MCMC) functionality}
\label{SecMCMC}

The Markov Chain Monte Carlo (MCMC) functionality used in {\it
StarSpotz} is fully described in \citet{Cro06a}. The goal is to take
an $n$-step intelligent random walk around the parameter space of
interest while recording the point in parameter space for each of the
$K$ fitted parameters for each step.  This samples the posterior
parameter distribution, which can be thought of in photometric spot
modeling as simply the range in each of the $K$ fitted parameters that provide
a reasonable fit to the light curve.  The advantage of a MCMC is that
it does not simply follow the path of steepest descent - resulting in
the possibility that it becomes stuck in a local minimum - but allows
the MCMC to fully explore the immediate parameter space.  The $K$
parameters can thus be defined by a $K$ dimensional vector {\bf y}.

T stands for the ``virtual temperature'' and
determines the probability that the MCMC chain will accept large deviations to higher regions
in $\chi^{2}$ space \citep{Sajina}.
The virtual temperature should properly be unity ($T=1$) to sample the posterior parameter
distribution when reduced $\chi^{2}$ is near unity. 
However, in photometric spot modeling, nature is often more complicated than one's model
can take into account, resulting in anomalously high reduced $\chi^{2}$ values.
Setting $T$ to be the
minimum reduced $\chi^{2}$ value observed
produces the same effect as scaling reduced $\chi^{2}$ to one, and thus
effectively samples the posterior parameter distribution. In this present application $T_{min}$
will refer to this value of $T$ equal to the minimum reduced $\chi^{2}$ observed.

A suitable burn-in period as described in \citet{Cro06a} is often used, or a $\chi^{2}$ cut where points with $\chi^{2}$ values
above this user-specified minimum are excised from the analysis. A user-specified thinning factor, $f$, \citep{Cro06a} is also used. 
To ensure proper convergence and mixing we employ the tests as proposed by \citet{Gelman} and explicitly
defined for our purposes in \citet{Cro06a}. We ensure the vector {\bf R} falls as close to 1.0 as possible
for each of the $K$ fitted parameters.
The marginalized likelihood (\citealt{Sajina}; \citealt{Lewis}) as described in $\S$3.1
of \citet{Cro06a} is used to define our 68\% credible regions, while the mean likelihood \citep{Sajina} is used for comparison. 
The mean likelihood is produced using the $\chi^{2}$ values in a particular bin
of a histogram of parameter values, while
the marginalized likelihood directly reflects the fraction of times that the MCMC chains are in a particular bin. Peaks in the marginalized
likelihood that are inconsistent with peaks in the mean likelihood 
can be indicative of the chains having yet to
converge, or larger multidimensional parameter space rather than
a better fit \citep{Lewis}. In our analysis
we quote the minimum to maximum range of the 68\% credible regions.

\subsection{Parallel Tempering}
\label{SecTemper}

\citet{Gregorypaper} applied parallel tempering to a radial velocity
search for extrasolar planets, as the parameter space he investigated
had widely seperated regions in parameter space that provided a
reasonable fit to the data.  In our application as we fit for a large
number of parameters, $K$=27, we use parallel tempering to explore our
multi-dimensional parameter space and search for other local minima in
$\chi^{2}$ space indicative of other plausible solutions, and thus
starspot configurations, that produce a reasonable fit to the
light-intensity curve.

In parallel tempering, $M_{P}$ multiple MCMC chains are run
simultaneously each with different values of the virtual temperature,
$T_{m}$, where $m$ is an index that runs from 1 to $M_{P}$.  Chains
with higher values of $T$ are easily able to jump to radically
different areas in parameter space ensuring that the chain does not
become stuck in a local minimum.  Chains with lower values of $T$ can
refine themselves and descend into local minima and possibly the global
minimum in $\chi^{2}$ space \citep{Gregorypaper}.  The chain with $T$ =
1.0$\times$$T_{min}$ is the target distribution and is referred to as
the ``cold sampler.'' The parameters of the ``cold sampler'' are used
for analysis to return the posterior parameter distribution.  A set
mean fraction of the time, $l$, a proposal is made for the parameters
{\bf y} of adjacent chains to be exchanged.  If this proposal is
accepted than adjacent chains are chosen at random and the $K$
parameters {\bf y} for each chain are exchanged.  In this way, this
method allows radically different areas in parameter space to be
investigated in the high temperature chains, before reaching the ``cold
sampler'' where a local minimum can be sought efficiently.  We run 11 parallel chains with virtual
temperatures, $T_{m}$, of:  $T$ = \{100$\times$$T_{min}$,
50.0$\times$$T_{min}$, 33.0$\times$$T_{min}$ , 20.0$\times$$T_{min}$,
10.0$\times$$T_{min}$, 5.0$\times$$T_{min}$, 3.00$\times$$T_{min}$,
2.00$\times$$T_{min}$, 1.50$\times$$T_{min}$ 1.25$\times$$T_{min}$,
1.00$\times$$T_{min}$  \}.

In a typical parallel tempering scheme the number of iterations between
swap proposals is set to some constant integer, $B$ (e.g. $B$ $\approx$
10).  With the large number of parameters being fitted across the three
epochs of data ($K$=27) in the current application, it was found that
this typical scheme was not able to efficiently reach the
low-$\chi^{2}$ valleys in the multi-dimensional parameter space
indicative of a reasonable fit to the light-intensity curve.
Thus the number of iterations between swap proposals, $B_{m}$, was set
to a variable number between the chains.  $B_{m}$ was set to a larger
number for the lower temperature chains to allow these chains a greater
number of iterations to refine themselves and explore the
low-$\chi^{2}$ areas.  This was found to be a suitable compromise as
the higher virtual temperature chains required fewer iterations to
reach radically different areas of parameter space, while the lower
virtual temperature chains required nearly an order of magnitude more
iterations to reach lower areas in $\chi^{2}$ space indicative of
reasonable fits to the light-curve.
The number of chain points per chain was set to B = \{$B_{min}$, $B_{min}$, $B_{min}$, $B_{min}$, $B_{min}$, 
$B_{min}$, $B_{min}$, 2.0$\times$$B_{min}$, 2.0$\times$$B_{min}$, 3.0$\times$$B_{min}$, 4.0$\times$$B_{min}$\}.
$B_{min}$ was set to twenty ($B_{min}$=20), within an order of magnitude of the value proposed by 
\citet{Gregorypaper}; extensive tests
have proven this choice to be reasonable.
$l$ was set as 0.80 to allow for exchanges between adjacent MCMC chains on average
80\% percent of the time.
That means a swap is approved if $u_{l}$ $<$ $l$, where $u_{l}$ is a random number ($u_{l} \ \epsilon [0,1]$).
Significant experimentation was needed to choose and refine suitable choices of $T_{m}$, $B_{m}$, $B_{min}$, and 
$l$.

In the present application we start 8$\times$11 of these
parallel tempering chains starting from random points in our acceptable parameter
space as given in the Parallel Tempering initial allowed range column of Table \ref{TablePrior}.
MCMC fitting was then implemented for $n$=4000 steps to ensure that the starting point for the parallel tempering chains provided
a mediocre fit to the light curve.
A mediocre fit was desired to ensure that the starting point for the parallel tempering chains
provided a reasonable fit to the light curve so as to reduce the required burn-in period, while not starting from within a
significant local minimum so as to allow the parallel tempering 
chains to efficiently explore the parameter space.
Each of these 8$\times$11 MCMC parallel tempering chains were then run
for $n_{x}$=13000 exchanges, resulting in $n_{x}$$\times$20$\times$4 = 1.04$\times$$10^{6}$ total steps for each of the 8 ``cold samplers.''
A burn-in period was not used in this application for our analysis. Rather a $\chi^{2}$ cut was used where all points of the parallel tempering ``cold samplers'' with reduced 
$\chi^{2}$ greater than some particular minimum, $\chi^{2}_{CUT}$, were excised from the analysis. This serves a similar function to a burn-in period
for the initial points, while allowing the analysis to focus on the chain points that provide a reasonable fit to the light curve. 
Given the large number of parameters we fit for, $K$=27, this is useful as the 
parallel tempering chains spend an inordinate amount of time exploring parameter spaces that do not provide a good fit to the data, and would otherwise
complicate the analysis. 
A thinning factor, $f$, of 20 were used in all chains for the parallel tempering analysis.
The 8$\times$11 parallel tempering chains required a total computational time of 7.0 CPU days on a Pentium processor with a clock speed
of 3.2 Ghz with 1.0 GB of memory.
This value of $n_{x}$ was chosen to ensure that the vector {\bf R}, in 
this case {\bf $R_{Parallel}$} (\citealt{Gelman}; \citealt{Cro06a}),
fell as close to 1.0 as possible for each of the $K=27$ parameters. 
{\bf R} $<$ $\approx$1.3 signifies that
the chain points have roughly converged and thus we are close to an equilibrium distribution.
To determine {\bf $R_{Parallel}$} we do not use the above $\chi^{2}$ cut but instead use a burn-in period of $n_{burn}$ = 30000 steps. This is because
we are interested in ensuring the parallel tempering chains have converged while exploring our parameter space as a whole, rather than the narrow region 
of parameter space signifying the global $\chi^{2}$ minimum.
Analysis of these 8 ``cold samplers'' is given below in $\S$\ref{SecTemperResults}.

\subsubsection{$\kappa$1 Ceti Parallel Tempering results}
\label{SecTemperResults}

The Parallel Tempering results as applied to {\it MOST}'s 2003-2005 observations of $\kappa^{1}$ Ceti indicated that 
there is a single unique solution to the data-set that provides the best fit.
Following the $n_{x}$=13000 exchanges as described above the vector {\bf $R_{Parallel}$} 
was investigated for each of the $K$=27 parameters.
It was found that {\bf $R_{Parallel}$} fell below 1.6 for all parameters, and {\bf $R_{Parallel}$} fell below 1.3 for most parameters.
These low values of {\bf $R_{Parallel}$} indicate adequate convergence has been obtained.
Each of the components of the vector {\bf $R_{Parallel}$} are given in Table \ref{TablePrior}.
These low values for all the $K$ parameters
of {\bf $R_{Parallel}$} indicate that our parallel tempering chains have approximately converged to an equilibrium, and adequately sampled the realistic
parameter space.

For our analysis we set $T_{min}$ as 2.96 as it is the global minimum observed in the below section ($\S$\ref{SecMCMCResults})
and a reduced $\chi^{2}_{CUT}$ of 3.38 as it was judged this was
reasonably close to the global $\chi^{2}$ minimum to focus the analysis exclusively
on the solutions that provided a reasonable fit to the light curve.
Our 68\% credible regions are given by the marginalized likelihood following the $\chi^{2}$ cut.
The immediate parameter space indicated by these results are given in Table \ref{TableMCMC}.
This parameter space was thus worthy of more detailed
investigation to explore this $\chi^{2}$ global minimum, as well as to
define best-fit values and appropriate uncertainties for the fitted and
derived parameters of interest. Also, possible
correlations between the fitted parameters were investigated for. 

Detailed investigation of the
parallel tempering results indicate that all other local minima produce
significantly worse fits to the data-sets.
The only minor exception indicated by the parallel tempering results is that
the second spot in the 2003 epoch can also be found at a latitude, $\beta_{2003\_2}$,
in the southern hemisphere as well as the northern hemisphere. Detailed investigation indicate
that the northern hemisphere solution
provides a significantly better fit and thus is the solution that will be investigated. 
Given the fact that all other local minima have been ruled out, and
the low values of {\bf $R_{Parallel}$} for all $K$ parameters, we
feel fully justified in stating that this parameter space is the unique
realistic solution to the observed light curve given our assumptions.

\begin{deluxetable*}{ccccc}
\tabletypesize{\footnotesize}
\tablecaption{$\kappa^{1}$ Ceti MCMC fitted parameters
\label{TablePrior}
}
% 
%  }
\tablewidth{0pt}
\tablehead{
\colhead{parameter}			&\colhead{prior allowed}					&\colhead{Parallel Tempering} 			&\colhead{ {\bf $R_{Parallel}$}  } 	&\colhead{ {\bf $R_{MCMC}$} }\\
\colhead{}				&\colhead{range}						&\colhead{initial allowed range}		&\colhead{   } 				&\colhead{  }\\
}
\startdata
$i$  ($^{o}$)					& 	$25$ - $85$, $30$ - $80$ \tablenotemark{a}	&	$25$ - $85$	 	& \RVALParallelTwoFour 	&\RVALTwoFour 	\\	
$k$  						& 	$-0.75$	- $0.75$				&	$-0.6$ - $0.6$		& \RVALParallelTwoFive 	&\RVALTwoFive\\	
$P_{EQ}$ (d)					&	$8.0$	-	$10.5$				&	$8.0$	-	$10.5$	& \RVALParallelTwoSix 	&\RVALTwoSix\\
$U_{2003}$					& 	0.99	-	1.01				&	1.000			& \RVALParallelSix 	&\RVALSix\\
$E_{2003\_1}$ (JD-2451545)			&	1409.0	-	1413.0				&	1409.0	-	1413.0	& \RVALParallelFour 	&\RVALFour\\
$\beta_{2003\_1}$ ($^{o}$)			&	$-90$	-	$90$				& $-10.0$ - $60.0$		& \RVALParallelTwo 	&\RVALTwo\\
$\gamma_{2003\_1}$ ($^{o}$)			&	$0.0$	-	$30$				& $4.0$ - $8.0$			& \RVALParallelZero 	&\RVALZero\\
$E_{2003\_2}$	(JD-2451545)			&	1404.0	-	1408.0				&	1404.0	-	1408.0	& \RVALParallelFive 	&\RVALFive\\
$\beta_{2003\_2}$ ($^{o}$)			&	$-90$	-	$90$				& $-10.0$ - $60.0$		& \RVALParallelThree 	&\RVALThree\\
$\gamma_{2003\_2}$ ($^{o}$)			&	$0.0$	-	$30$				& $4.0$ - $8.0$			& \RVALParallelOne 	&\RVALOne\\
$U_{2004}$					& 	0.99	-	1.015				&	1.000			& \RVALParallelOneSix 	&\RVALOneSix\\
$E_{2004\_1}$ (JD-2451545)			&	1754.7	-	1758.7				&	1754.7	-	1758.7	& \RVALParallelOneThree	&\RVALOneThree\\
$\beta_{2004\_1}$ ($^{o}$)			&	$-90$	-	$90$				& $-10.0$ - $60.0$		& \RVALParallelOneZero 	&\RVALOneZero\\
$\gamma_{2004\_1}$ ($^{o}$)			&	$0.0$	-	$30$				& $2.5$ - $6.0$			& \RVALParallelSeven 	&\RVALSeven\\
$E_{2004\_2}$	(JD-2451545)			&	1761.5	-	1765.5				&	1761.5	-	1765.5	& \RVALParallelOneFour 	&\RVALOneFour\\
$\beta_{2004\_2}$ ($^{o}$)			&	$-90$	-	$90$				& $-10.0$ - $60.0$		& \RVALParallelOneOne 	&\RVALOneOne\\
$\gamma_{2004\_2}$ ($^{o}$)			&	$0.0$	-	$30$				& $2.5$ - $6.0$			& \RVALParallelEight 	&\RVALEight\\
$E_{2004\_3}$	(JD-2451545)			&	1767.1	-	1771.1				& 1767.1	-	1771.1	& \RVALParallelOneFive 	&\RVALOneFive\\
$\beta_{2004\_3}$ ($^{o}$)			&	$-90$	-	$90$				& $-10.0$ - $60.0$		& \RVALParallelOneTwo 	&\RVALOneTwo\\
$\gamma_{2004\_3}$ ($^{o}$)			&	$0.0$	-	$30$				& $2.5$ - $6.0$			& \RVALParallelNine 	&\RVALNine\\
$U_{2005}$					& 	0.99	-	1.01				&	1.000			& \RVALParallelTwoThree	&\RVALTwoThree\\
$E_{2005\_1}$ (JD-2451545)			&	2115.0	-	2119.0				& 2115.0	-	2119.0	& \RVALParallelTwoOne 	&\RVALTwoOne\\
$\beta_{2005\_1}$ ($^{o}$)			&	$-90$	-	$90$				& $-10.0$ - $60.0$		& \RVALParallelOneNine 	&\RVALOneNine\\
$\gamma_{2005\_1}$ ($^{o}$)			&	$0.0$	-	$30$				& $4.0$ - $8.0$			& \RVALParallelOneSeven &\RVALOneSeven\\
$E_{2005\_2}$	(JD-2451545)			&	2118.1	-	2122.1				& 2118.1	-	2122.1	& \RVALParallelTwoTwo 	&\RVALTwoTwo\\
$\beta_{2005\_2}$ ($^{o}$)			&	$-90$	-	$90$				& $-10.0$ - $60.0$		& \RVALParallelTwoZero 	&\RVALTwoZero\\
$\gamma_{2005\_2}$ ($^{o}$)			&	$0.0$	-	$30$				& $4.0$ - $8.0$			& \RVALParallelOneEight &\RVALOneEight\\

\enddata
\tablenotetext{a}{The left hand values were used for the parallel tempering application ($\S$\ref{SecTemperResults}), while the values on the right were used
for the MCMC application ($\S$\ref{SecMCMCResults}).}
\end{deluxetable*}

\subsection{$\kappa^1$  Ceti MCMC application}
\label{SecMCMCResults}

In $\S$\ref{SecTemper} we demonstrated that the solution presented is
unique for the 2003-2005 {\it MOST} epochs under the $k$, $P_{EQ}$,
paradigm.  Thus we used the MCMC methods discussed in \citet{Cro06a} and
summarized above in $\S$\ref{SecMCMC} to investigate possible
correlations and derive best-fit values and uncertainties in the fitted
and derived parameters for $\kappa^{1}$ Ceti.  We used $M$=4 chains
starting from reasonably different points in parameter space, although
within the local $\chi^{2}$ minimum that defined the parameter space of interest of
$\S$\ref{SecTemper}. The prior ranges are given in Table
\ref{TablePrior}. The choices in prior are identical to those used for
the Parallel Tempering section except a slightly more conservative
range is used for the inclination angle, 30$^{o}$ $<$ $i$  $<$
80$^{o}$, to limit the MCMC chain to a realistic parameter space.  We
run each of the $M$=4 chains for $n$=7.0$\times$$10^{6}$ steps, requiring
a total computational time of 8.0 CPU days on the aforementioned processor.
This
large value of $n$ was motivated to ensure that the {\bf R} vector, in this case {\bf $R_{MCMC}$},
fell below 1.20 for all $K$=27 parameters, and fell below 1.05 for most parameters, indicative of suitable convergence.
The values
of each of the $K$ components of {\bf $R_{MCMC}$} are given in Table
\ref{TablePrior}.
 
The results of this analysis, henceforth referred to as Solution 1, 
can be seen in Figures 2 and 3, and are summarized in 
Table \ref{TableMCMC}. As
one might expect, especially given the similar results of $\epsilon$
Eri \citep{Cro06a}, the latitudes of the various spots are
moderately correlated with the inclination angle of the system, $i$.
However, the differential rotation coefficient, $k$, is largely
uncorrelated with $i$. This is in stark contrast to the results of $\epsilon$
Eri where the correlation of spot latitude with $i$ produced a moderate
correlation between $k$ and $i$.  Our derived value of $k$ = \kLow \ -
\kHigh \ is therefore very robust as it is does not require an
independent estimate of $i$.

It is important to note, though, that the 2003 and 2005 {\it MOST}
data-sets do not severely limit the allowed best-fit range of the
differential rotation coefficient, $k$.  The 2003 and 2005 {\it MOST}
data-sets were best-fit with spots that were not widely seperated in
latitude, and thus these epochs contributed only marginally to the determination
of the differential rotation coefficient.  The 2004 {\it MOST} data-set
most severely limited the best-fit range of the differential rotation
coefficient, as it required three spots ranging in latitude from the mid-southern hemisphere,
to the equator, to near the north pole.  

Also, although we have used a relatively liberal flat prior on the
stellar inclination angle, $i$, of 30$^{o}$ $<$ $i$ $<$ 80$^{o}$, the
value of $i$ returned from our fit gives $v$sin$i$ = \vsiniLow \ - \vsiniHigh \ km s$^{-1}$, entirely consistent with the
spectroscopic value of $v$sin$i$ = 4.64 $\pm$ 0.11 km s$^{-1}$
determined by \citet{Ruc04} from the width of the Ca II K-line
reversals. This agreement is a very useful sanity check and gives us
great confidence that our fitted results and our assumptions of
circular spots are in close agreement with the actual behaviour of the
star during these three epochs.

The minimum $\chi^{2}$ point observed in our MCMC chains is also quoted in Table \ref{TableMCMC}. The fit to the light curves in 
2003, 2004, 2005 and the views of the modelled spots shown from the line of sight (LOS) of this minimum $\chi^{2}$ solution 
are given in Figures \ref{Fig2003}, \ref{Fig2004}, and \ref{Fig2005}.

Our current results are largely consistent with the analysis of the
2003 {\it MOST} data by \citet{Ruc04} as can be seen in the comparison
given in Table \ref{TableMCMC}. The only discrepancies of note are that
our MCMC analysis indicates a slightly smaller value of the inclination
angle, $i$ = $\iLow \ -  \iHigh$$^{o}$, and that we find a shorter
period of the 2$^{nd}$ spot, $p_{2003\_2}$ = $\pOThreeTWOLow \ -
\pOThreeTWOHigh$ d. The period found by \citet{Ruc04} for the 2$^{nd}$
spot depends on effective removal of the variations caused by the
larger spot  and it was not possible to asssign a reliable latitude.
They also asumed that the spots were black making them
smaller than those in this paper.

\clearpage
\begin{deluxetable*}{cccccc}
\tabletypesize{\footnotesize}
\tablecaption{$\kappa^1$ Ceti MCMC fitted parameters
\label{TableMCMC}
}
\tablewidth{0pt}
\tablehead{
	\colhead{parameter}				&\colhead{fitted}	&\colhead{Rucinski}				&\colhead{Parallel Tempering}				&\colhead{Minimum $\chi^{2}$}		&\colhead{MCMC}			\\
	\colhead{}					&\colhead{}		&\colhead{2004 results}					&\colhead{results}					&\colhead{solution}			&\colhead{{\bf Solution 1}}		\\
}
\startdata
Reduced $\chi^{2}$					& n/a			& - 							&	\chiMeLow \ - \chiMeHigh			&	\chimin				&	\chiLow - \chiHigh			\\	
$\nu$	\tablenotemark{a}				& n/a			& -							&	817						&	817				&	817					\\
$i$  ($^{o}$)						& yes			&$70 \pm 4$						&	\iMeLow \ - \iMeHigh				& 	\imin				&	\iLow \ - \iHigh			\\	
$k$  							& yes			& - 							&	\kMeLow \ - \kMeHigh				& 	\kmin				&	\kLow \ - \kHigh			\\	
$P_{EQ}$						& yes			& - 							&	\PEQMeLow \ - \PEQMeHigh			&	\PEQmin				&	\PEQLow \ - \PEQHigh			\\
$v$sin$i$ (km s$^{-1}$)	\tablenotemark{c}		& derived		& 4.64 $\pm$ 0.11 \tablenotemark{b}			&	\vsiniMeLow \ - \vsiniMeHigh			& 	\vsinimin			&	\vsiniLow \ - \vsiniHigh		\\
equatorial speed (km s$^{-1}$)	\tablenotemark{c}	& derived		& -							&	\eqspeedMeLow \ - \eqspeedMeHigh		&	\eqspeedmin			&	\eqspeedLow \ - \eqspeedHigh		\\
$u$							& assumed		& 0.80							&	0.6840						&	0.684				&	0.684	\\
$\kappa_{\omega}$					& assumed		& 0.00							&	0.220 						&	0.220				&	0.220	\\
$U_{2003}$						& yes			& - 							&	\UOThreeMeLow \ - \UOThreeMeHigh		& 	\UOThreemin			&	\UOThreeLow \ - \UOThreeHigh		\\
$E_{2003\_1}$ (JD-2451545)				& yes			& - 							&	\EOThreeONEMeLow \ - \EOThreeONEMeHigh		&	\EOThreeONEmin			&	\EOThreeONELow \ - \EOThreeONEHigh	\\
$p_{2003\_1}$ (days)					& derived		&$8.9 \pm 0.1$	\tablenotemark{d}			&	\pOThreeONEMeLow \ - \pOThreeONEMeHigh		&	\pOThreeONEmin			&	\pOThreeONELow \ - \pOThreeONEHigh	\\
$\beta_{2003\_1}$ ($^{o}$)				& yes			&$40 \pm 7$						&	\latOThreeONEMeLow \ - \latOThreeONEMeHigh	&	\latOThreeONEmin		&	\latOThreeONELow \ - \latOThreeONEHigh	\\
$\gamma_{2003\_1}$ ($^{o}$)				& yes			&$11.1 \pm 0.6$						&	\gOThreeONEMeLow \ - \gOThreeONEMeHigh		&	\gOThreeONEmin			&	\gOThreeONELow \ - \gOThreeONEHigh	\\
$E_{2003\_2}$	(JD-2451545)				& yes			& - 							&	\EOThreeTWOMeLow \ - \EOThreeTWOMeHigh		&	\EOThreeTWOmin			&	\EOThreeTWOLow \ - \EOThreeTWOHigh	\\
$p_{2003\_2}$ (days)					& derived		&$9.3 - 9.7$	\tablenotemark{d}			&	\pOThreeTWOMeLow \ - \pOThreeTWOMeHigh		&	\pOThreeTWOmin			&	\pOThreeTWOLow \ - \pOThreeTWOHigh	\\
$\beta_{2003\_2}$ ($^{o}$)				& yes			& -							&	\latOThreeTWOMeLow \ - \latOThreeTWOMeHigh	&	\latOThreeTWOmin		&	\latOThreeTWOLow \ - \latOThreeTWOHigh	\\
$\gamma_{2003\_2}$ ($^{o}$)				& yes			& -							&	\gOThreeTWOMeLow \ - \gOThreeTWOMeHigh		&	\gOThreeTWOmin			&	\gOThreeTWOLow \ - \gOThreeTWOHigh	\\
$U_{2004}$						& yes			&n/a							&	\UOFourMeLow \ - \UOFourMeHigh			& 	\UOFourmin			&	\UOFourLow \ - \UOFourHigh		\\
$E_{2004\_1}$ (JD-2451545)				& yes			&n/a							&	\EOFourONEMeLow \ - \EOFourONEMeHigh		&	\EOFourONEmin			&	\EOFourONELow \ - \EOFourONEHigh	\\
$p_{2004\_1}$ (days)					& derived		&n/a							&	\pOFourONEMeLow \ - \pOFourONEMeHigh		&	\pOFourONEmin			&	\pOFourONELow \ - \pOFourONEHigh	\\
$\beta_{2004\_1}$ ($^{o}$)				& yes			&n/a							&	\latOFourONEMeLow \ - \latOFourONEMeHigh	&	\latOFourONEmin			&	\latOFourONELow \ - \latOFourONEHigh	\\
$\gamma_{2004\_1}$ ($^{o}$)				& yes			&n/a							&	\gOFourONEMeLow \ - \gOFourONEMeHigh		&	\gOFourONEmin			&	\gOFourONELow \ - \gOFourONEHigh	\\
$E_{2004\_2}$	(JD-2451545)				& yes			&n/a							&	\EOFourTWOMeLow \ - \EOFourTWOMeHigh		&	\EOFourTWOmin			&	\EOFourTWOLow \ - \EOFourTWOHigh	\\
$p_{2004\_2}$ (days)					& derived		&n/a							&	\pOFourTWOMeLow \ - \pOFourTWOMeHigh		&	\pOFourTWOmin			&	\pOFourTWOLow \ - \pOFourTWOHigh	\\
$\beta_{2004\_2}$ ($^{o}$)				& yes			&n/a							&	\latOFourTWOMeLow \ - \latOFourTWOMeHigh	&	\latOFourTWOmin			&	\latOFourTWOLow \ - \latOFourTWOHigh	\\
$\gamma_{2004\_2}$ ($^{o}$)				& yes			&n/a							&	\gOFourTWOMeLow \ - \gOFourTWOMeHigh		&	\gOFourTWOmin			&	\gOFourTWOLow \ - \gOFourTWOHigh	\\
$E_{2004\_3}$	(JD-2451545)				& yes			&n/a							&	\EOFourTHREEMeLow \ - \EOFourTHREEMeHigh	&	\EOFourTHREEmin			&	\EOFourTHREELow \ - \EOFourTHREEHigh	\\
$p_{2004\_3}$ (days)					& derived		&n/a							&	\pOFourTHREEMeLow \ - \pOFourTHREEMeHigh	&	\pOFourTHREEmin			&	\pOFourTHREELow \ - \pOFourTHREEHigh	\\
$\beta_{2004\_3}$ ($^{o}$)				& yes			&n/a							&	\latOFourTHREEMeLow \ - \latOFourTHREEMeHigh	&	\latOFourTHREEmin		&	\latOFourTHREELow \ - \latOFourTHREEHigh\\
$\gamma_{2004\_3}$ ($^{o}$)				& yes			&n/a							&	\gOFourTHREEMeLow \ - \gOFourTHREEMeHigh	&	\gOFourTHREEmin			&	\gOFourTHREELow \ - \gOFourTHREEHigh	\\
$U_{2005}$						& yes			&n/a							&	\UOFiveMeLow \ - \UOFiveMeHigh			& 	\UOFivemin			&	\UOFiveLow \ - \UOFiveHigh		\\
$E_{2005\_1}$ (JD-2451545)				& yes			&n/a							&	\EOFiveONEMeLow \ - \EOFiveONEMeHigh		&	\EOFiveONEmin			&	\EOFiveONELow \ - \EOFiveONEHigh	\\
$p_{2005\_1}$ (days)					& derived		&n/a							&	\pOFiveONEMeLow \ - \pOFiveONEMeHigh		&	\pOFiveONEmin			&	\pOFiveONELow \ - \pOFiveONEHigh	\\
$\beta_{2005\_1}$ ($^{o}$)				& yes			&n/a							&	\latOFiveONEMeLow \ - \latOFiveONEMeHigh	&	\latOFiveONEmin			&	\latOFiveONELow \ - \latOFiveONEHigh	\\
$\gamma_{2005\_1}$ ($^{o}$)				& yes			&n/a							&	\gOFiveONEMeLow \ - \gOFiveONEMeHigh		&	\gOFiveONEmin			&	\gOFiveONELow \ - \gOFiveONEHigh	\\
$E_{2005\_2}$	(JD-2451545)				& yes			&n/a							&	\EOFiveTWOMeLow \ - \EOFiveTWOMeHigh		&	\EOFiveTWOmin			&	\EOFiveTWOLow \ - \EOFiveTWOHigh	\\
$p_{2005\_2}$ (days)					& derived		&n/a							&	\pOFiveTWOMeLow \ - \pOFiveTWOMeHigh		&	\pOFiveTWOmin			&	\pOFiveTWOLow \ - \pOFiveTWOHigh	\\
$\beta_{2005\_2}$ ($^{o}$)				& yes			&n/a							&	\latOFiveTWOMeLow \ - \latOFiveTWOMeHigh	&	\latOFiveTWOmin			&	\latOFiveTWOLow \ - \latOFiveTWOHigh	\\
$\gamma_{2005\_2}$ ($^{o}$)				& yes			&n/a							&	\gOFiveTWOMeLow \ - \gOFiveTWOMeHigh		&	\gOFiveTWOmin			&	\gOFiveTWOLow \ - \gOFiveTWOHigh	\\
\enddata
\tablenotetext{a}{$\nu$ is the number of binned data points minus the number of fitted parameters.}
\tablenotetext{b}{spectroscopic value from \citet{Ruc04}, not derived photometrically.}
\tablenotetext{c}{These values determined using $R_{*}$ = 0.95 $R_{\bigodot}$.}
\tablenotetext{d}{This parameter fitted, rather than derived \citep{Ruc04}.}

\end{deluxetable*}

\subsection{$\kappa^1$  Ceti 2004 Parallel Tempering and MCMC application}
\label{SecTwoThousandFourResults}

Given the good agreement of the {\it MOST} 2003, 2004 and 2005
$\kappa^{1}$ Ceti observations with the assumed solar-type differential
rotation profile in Equation 1 and summarized in Figure
\ref{FigPversusB}, we decided to test independently whether solar-type
differential rotation is indeed present in $\kappa^{1}$ Ceti.  The 2004
light curve was chosen because, as noted above, it constrained the
equatorial period, $P_{EQ}$, and differential rotation coefficient,
$k$, more significantly than the light curves from the other two
years.

For this independent test we fitted explicitly for the periods of each
of the three spots: $p_{2004\_1}$, $p_{2004\_2}$, and $p_{2004\_3}$.
The $K$=13 fitted parameters are summarized in Table
\ref{TableTwoThousandFourMCMC}. For these we used the same priors, and
for the parallel tempering chains we started from the same random
ranges in parameter space, as summarized in Table \ref{TablePrior}. We
assumed $i$= \imin$^{o}$ corresponding to the minimum $\chi^{2}$ value
found above. This assumption for the inclination angle is justified
because it results in $v$sin$i$ = \vsinimin km s$^{-1}$(assuming
$P_{EQ}$ $\approx$ 8.78), a value close to the spectroscopic value
measured by \citet{Ruc04}.

Parallel tempering as described in $\S$\ref{SecTemper} was used to find
the unique global minimum. Starting from random points in acceptable
parameter space we implemented MCMC fitting for $n$=3000 steps
to ensure a mediocre fit to
the light curve. Parallel tempering chains were then implemented with
$n_{x}$=1800 exchanges resulting in 144000 steps for each of the 8
``cold samplers''.  We used a burn-in period of the first 5000
steps to determine the {\bf R} vector, in this case {\bf
$R_{Parallel}^{2004}$}. All $K$=13 parameters of {\bf
$R_{Parallel}^{2004}$} fell below 1.4, and below 1.2 for most, as
summarized in Table \ref{TableTwoThousandFourMCMC}. Analysis is
performed using a $T_{min}$ of 1.50 and a reduced $\chi^{2}_{CUT}$ of
2.0. The parallel tempering results identify a unique global minimum as
summarized in Table \ref{TableTwoThousandFourMCMC}, that is nearly
identical to Solution 1 as summarized above in $\S$\ref{SecMCMCResults}
and Table 3.

Thus after using parallel tempering to identify the unique global
minimum, we explored this parameter space with our MCMC techniques as
described above in $\S$\ref{SecMCMC} and $\S$\ref{SecMCMCResults}. We
use $n$=3.7$\times$$10^{6}$ steps, and then use a burn-in period of
1000 steps to determine the {\bf R} vector (in this case {\bf
$R_{MCMC}^{2004}$}).  The parameters of {\bf $R_{MCMC}^{2004}$} fell
below 1.02 for all $K$=13 parameters as summarized in Table
\ref{TableTwoThousandFourMCMC}, and thus demonstrated more than adequate convergence.
Our MCMC results are also summarized in
Table \ref{TableTwoThousandFourMCMC}.

As summarized in the bottom-panel of Figure \ref{FigPversusB} these
results indicate that the 2004 {\it MOST} data-set provides a strong
independent argument that solar-type differential rotation pattern
defined by Equation 1 has been observed in $\kappa^{1}$ Ceti.  Using
only the 2004 data-set the values of the differential rotation
coefficient, and equatorial period are:  $k$ = \kFourLow \ -
\kFourHigh, and $P_{EQ}$ = \PEQFourLow \ - \PEQFourHigh $d$.  These
values were determined by averaging for each point of the MCMC chains
the three possible $k$ and $P_{EQ}$ values, between spots 1 \& 2, 1 \&
3, and 2 \& 3.

\begin{deluxetable*}{cccccc}
\tabletypesize{\scriptsize}
\tablecaption{$\kappa^1$ Ceti 2004 MCMC fitted parameters
\label{TableTwoThousandFourMCMC}
}
\tablewidth{0pt}
\tablehead{
	\colhead{parameter}			&\colhead{fitted}	&\colhead{\bf $R_{Parallel}^{2004}$}	&\colhead{Parallel Tempering}					&\colhead{\bf $R_{MCMC}^{2004}$}	&\colhead{MCMC}						\\
	\colhead{}				&\colhead{}		&					&\colhead{results}						&					&\colhead{Solution}				\\
}
\startdata
Reduced $\chi^{2}$				& n/a			&	n/a				&	\chiMeFourLow \ - \chiMeFourHigh			&	n/a			&	\chiFourLow \ - \chiFourHigh			\\	
$\nu$	\tablenotemark{a}			& n/a			&	n/a				&	269							&	n/a			&	269						\\
$i$  ($^{o}$)					& assumed		&	n/a				&	\imin							&	n/a			&	\imin						\\	
$k$  						& derived		&	n/a				&	\kMeFourLow \ - \kMeFourHigh				&	n/a			&	\kFourLow \ - \kFourHigh			\\	
$v$sin$i$ (km s$^{-1}$)	\tablenotemark{b}	& derived		&	n/a				&	\vsiniMeFourLow \ - \vsiniMeFourHigh			&	n/a			&	\vsiniFourLow \ - \vsiniFourHigh		\\
$P_{EQ}$					& derived		&	n/a				&	\PEQMeFourLow \ - \PEQMeFourHigh			&	n/a			&	\PEQFourLow \ - \PEQFourHigh			\\
$u$						& assumed		&	n/a				&	0.6840							&	n/a			&	0.684						\\
$\kappa_{\omega}$				& assumed		&	n/a				&	0.220 							&	n/a			&	0.220						\\
$U_{2004}$					& yes			&	\RVALFourParallelOneTwo		&	\UOFourMeFourLow \ - \UOFourMeFourHigh			&	\RVALFourOneTwo		&	\UOFourFourLow \ - \UOFourFourHigh		\\
$E_{2004\_1}$ (JD-2451545)			& yes			&	\RVALFourParallelSix		&	\EOFourONEMeFourLow \ - \EOFourONEMeFourHigh		&	\RVALFourSix		&	\EOFourONEFourLow \ - \EOFourONEFourHigh	\\
$p_{2004\_1}$ (days)				& yes			&	\RVALFourParallelNine		&	\pOFourONEMeFourLow \ - \pOFourONEMeFourHigh		&	\RVALFourNine		&	\pOFourONEFourLow \ - \pOFourONEFourHigh	\\
$\beta_{2004\_1}$ ($^{o}$)			& yes			&	\RVALFourParallelThree		&	\latOFourONEMeFourLow \ - \latOFourONEMeFourHigh	&	\RVALFourThree		&	\latOFourONEFourLow \ - \latOFourONEFourHigh	\\
$\gamma_{2004\_1}$ ($^{o}$)			& yes			&	\RVALFourParallelZero		&	\gOFourONEMeFourLow \ - \gOFourONEMeFourHigh		&	\RVALFourZero		&	\gOFourONEFourLow \ - \gOFourONEFourHigh	\\
$E_{2004\_2}$	(JD-2451545)			& yes			&	\RVALFourParallelSeven		&	\EOFourTWOMeFourLow \ - \EOFourTWOMeFourHigh		&	\RVALFourSeven		&	\EOFourTWOFourLow \ - \EOFourTWOFourHigh	\\
$p_{2004\_2}$ (days)				& yes			&	\RVALFourParallelOneZero	&	\pOFourTWOMeFourLow \ - \pOFourTWOMeFourHigh		&	\RVALFourOneZero	&	\pOFourTWOFourLow \ - \pOFourTWOFourHigh	\\
$\beta_{2004\_2}$ ($^{o}$)			& yes			&	\RVALFourParallelFour		&	\latOFourTWOMeFourLow \ - \latOFourTWOMeFourHigh	&	\RVALFourFour		&	\latOFourTWOFourLow \ - \latOFourTWOFourHigh	\\
$\gamma_{2004\_2}$ ($^{o}$)			& yes			&	\RVALFourParallelOne		&	\gOFourTWOMeFourLow \ - \gOFourTWOMeFourHigh		&	\RVALFourOne		&	\gOFourTWOFourLow \ - \gOFourTWOFourHigh	\\
$E_{2004\_3}$	(JD-2451545)			& yes			&	\RVALFourParallelEight		&	\EOFourTHREEMeFourLow \ - \EOFourTHREEMeFourHigh	&	\RVALFourEight		&	\EOFourTHREEFourLow \ - \EOFourTHREEFourHigh	\\
$p_{2004\_3}$ (days)				& yes			&	\RVALFourParallelOneOne		&	\pOFourTHREEMeFourLow \ - \pOFourTHREEMeFourHigh	&	\RVALFourOneOne		&	\pOFourTHREEFourLow \ - \pOFourTHREEFourHigh	\\
$\beta_{2004\_3}$ ($^{o}$)			& yes			&	\RVALFourParallelFive		&	\latOFourTHREEMeFourLow \ - \latOFourTHREEMeFourHigh	&	\RVALFourFive		&	\latOFourTHREEFourLow \ - \latOFourTHREEFourHigh\\
$\gamma_{2004\_3}$ ($^{o}$)			& yes			&	\RVALFourParallelNine		&	\gOFourTHREEMeFourLow \ - \gOFourTHREEMeFourHigh	&	\RVALFourNine		&	\gOFourTHREEFourLow \ - \gOFourTHREEFourHigh	\\
\enddata
\tablenotetext{a}{$\nu$ is the number of binned data points minus the number of fitted parameters.}
\tablenotetext{b}{These values determined using $R_{*}$ = 0.95 $R_{\bigodot}$.}
\end{deluxetable*}

\section{Discussion}

The MCMC Solution 1 in Table 3 provides our best values for the various
spot and stellar parameters. The solutions are expressed by the 68\%
credible ranges rather than best values with standard deviations since
the likelihood histograms are non-gaussian. Figure \ref{FigPversusB} is
a good summary of our analysis. The periods and $|\beta|$ 68\%
marginalized contours are shown for each of the spots together with
mean and limiting curves for $k$ (\kBest, \kLow, \kHigh) and the range of
$P_{EQ}$ (\PEQBest, \PEQLow, \PEQHigh$d$). Each of the three years is distinguished by a
different color. $|\beta|$ ranges from 10$^{\circ}$ and 75$^{\circ}$
with the 2004 data obviously providing the most rigorous constraint on
$k$. Indeed, it demonstrates that \citet{Ruc04} had a most challenging
task in demonstrating differential rotation because the spots in 2003 were at
very similar latitudes with the second spot being particularly small. 

The analysis defined $k$ using Equation 1 which is derived from the solar pattern.
The agreement of all seven spots with the form of the $k$ curve as well
as the good agreement of the spot solutions in Figures 5, 6 and 7 suggests
that the the differential rotation curve for $\kappa^1$~Cet is closely similar to solar. The analysis of the 2004 light curve independently of any assumption about $k$ offers strong confirmation that the pattern is indeed solar.

For the Sun, $k$ has been derived quite independently from either sunspot latitudes
and periods \citep{New51} or surface radial velocities \citep{How83} yielding
$k$ = 0.19 and 0.12, respctively \citep{Kit05}.  The large discrepancy between these values may be
related in part to the different behavior of sunspot and photospheric motions.
In our case, we  depend on spots to determine $k$ and our value of 0.09  is
significantly lower than either of those for the Sun. This is in line with
calculations by \citet{Bro04} who find that $k$ should increase with age.

\citet{Gud97} estimated an age of 750 Myr for $\kappa^1$~Cet from their estimated 9.2 d rotation period. Our value of $P_{EQ}$ = \PEQBest \ $d$
suggests a still younger age. 

All of the photometric periods found to date for $\kappa^1$~Cet can
be explained by spots appearing at different latitudes. A period  of 9.09 d  is given
in the  {\it Hipparcos\/} catalog \citep{hip} while \citet{MG2002} quote a value of 9.214 d. \citet{Baliu1995} monitored Ca II H \& K photoelectrically between 1967 and 1991 and found
a rotational period of $9.4 \pm 0.1$ d \citep{Bal83} and \citet{Shk03} found a closely similar
period of $\sim$9.3 d mostly from spectra taken in 2002. While the apparent persistance of
this period which corresponds to a range of 50$^{\circ}$ to 60$^{\circ}$ in latitude for some
35 years maybe fortuitous, it might be related to some large scale magnetic structure.

\acknowledgments 
The Natural Sciences and Engineering Research Council of Canada supports the research of B.C., D.B.G., J.M.M., A.F.J.M., S.M.R.,  G.A.H.W.. Additional support for A.F.J.M. comes from FCAR (Qu\'ebec). R.K. is supported by the Canadian Space Agency. W.W.W. is supported by the Austrian Space Agency and the Austrian Science Fund (P14984).

\clearpage

%% the file, uncomment the \plotone, \plottwo, and

\begin{figure}
\epsscale{0.97}
\plotone{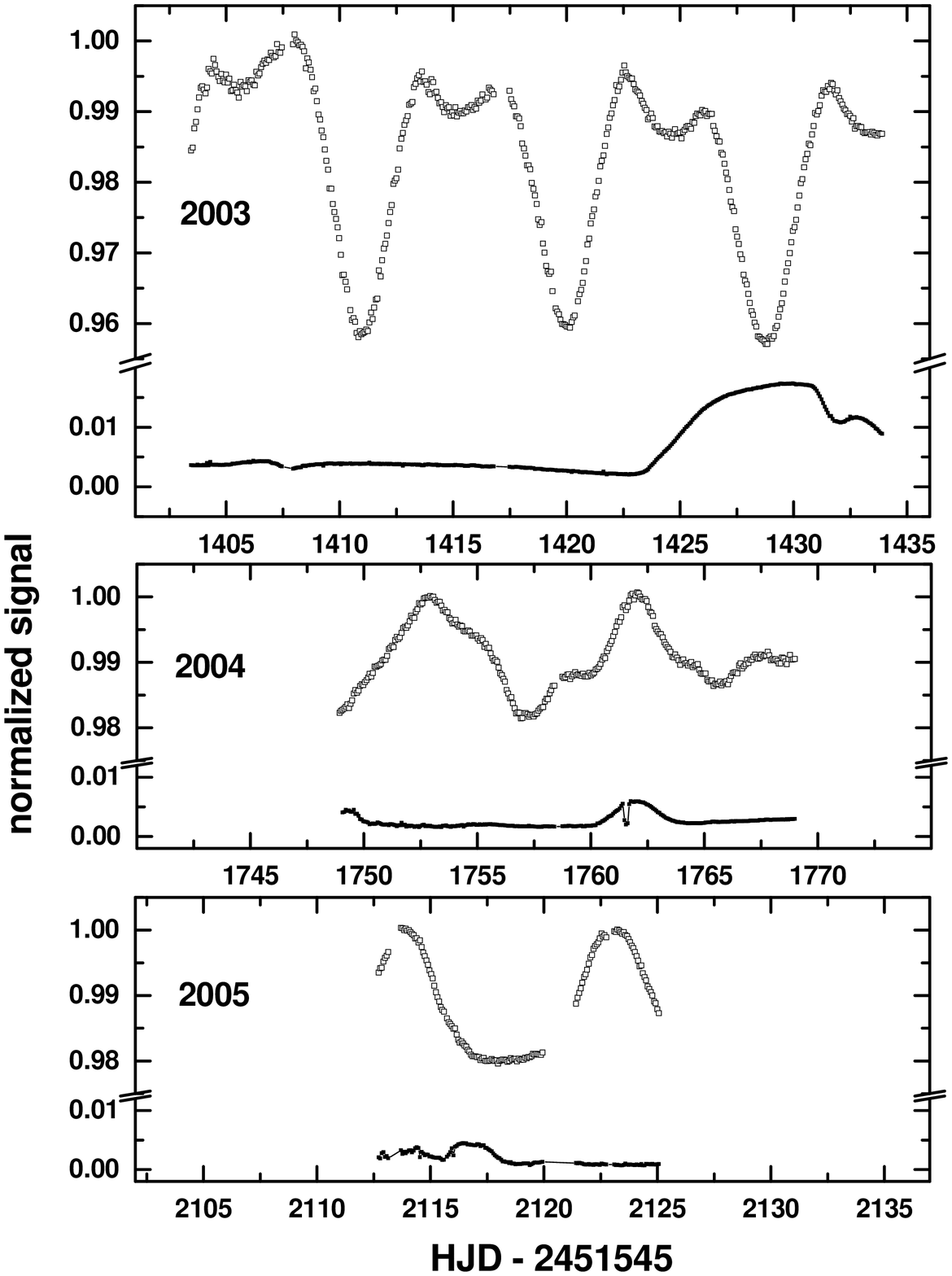}
\caption{{{\it MOST} light curves for $\kappa^{1}$ Ceti from 2003, 2004, and 2005. The points are the mean signals from individual 101.413 min satellite orbits. The black connected symbols in the lower part of each panel are the the differences between the highest and lowest of 7 adjacent simultaneously recorded (background) Fabry images (see text). The 'dip' at 1762 in 2004 corresponds to a total eclipse of the Moon. }  
	\label{FigRainer}
}
\end{figure}

\begin{figure}
\epsscale{1.0}
\includegraphics[width=1.55in, height = 3.0in, angle=270]{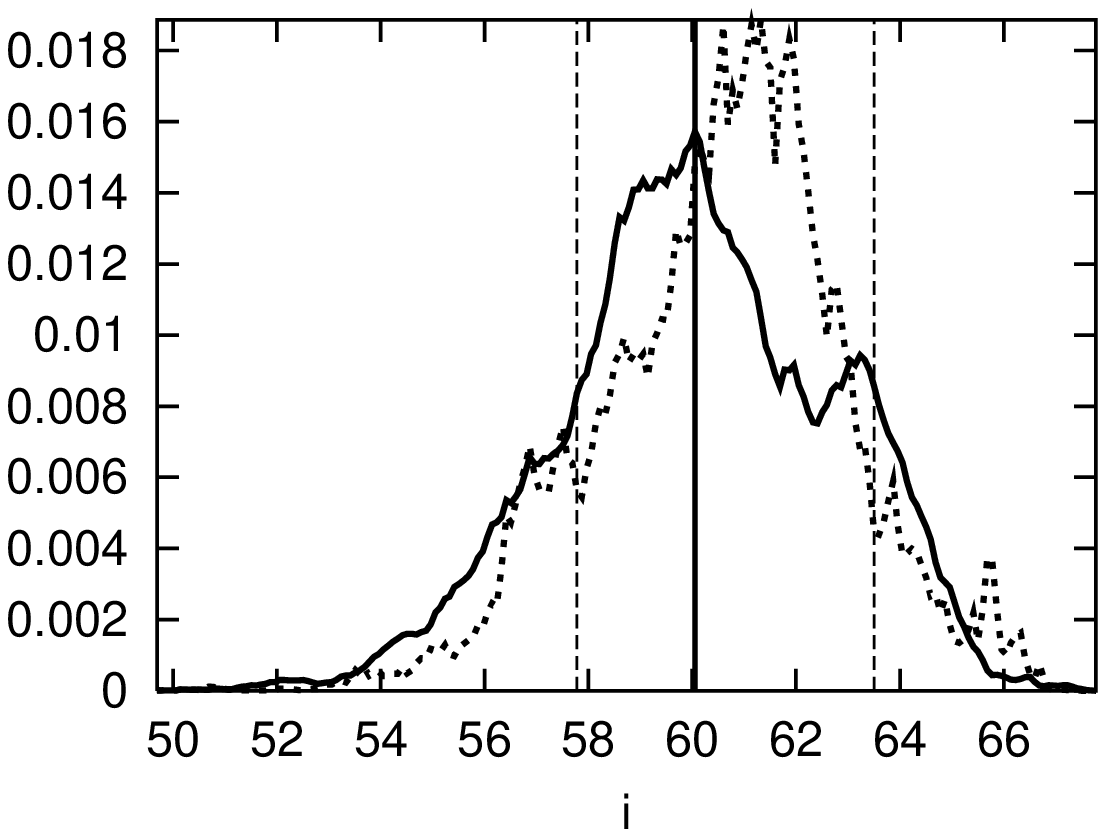}
\includegraphics[width=1.55in, height = 3.0in, angle=270]{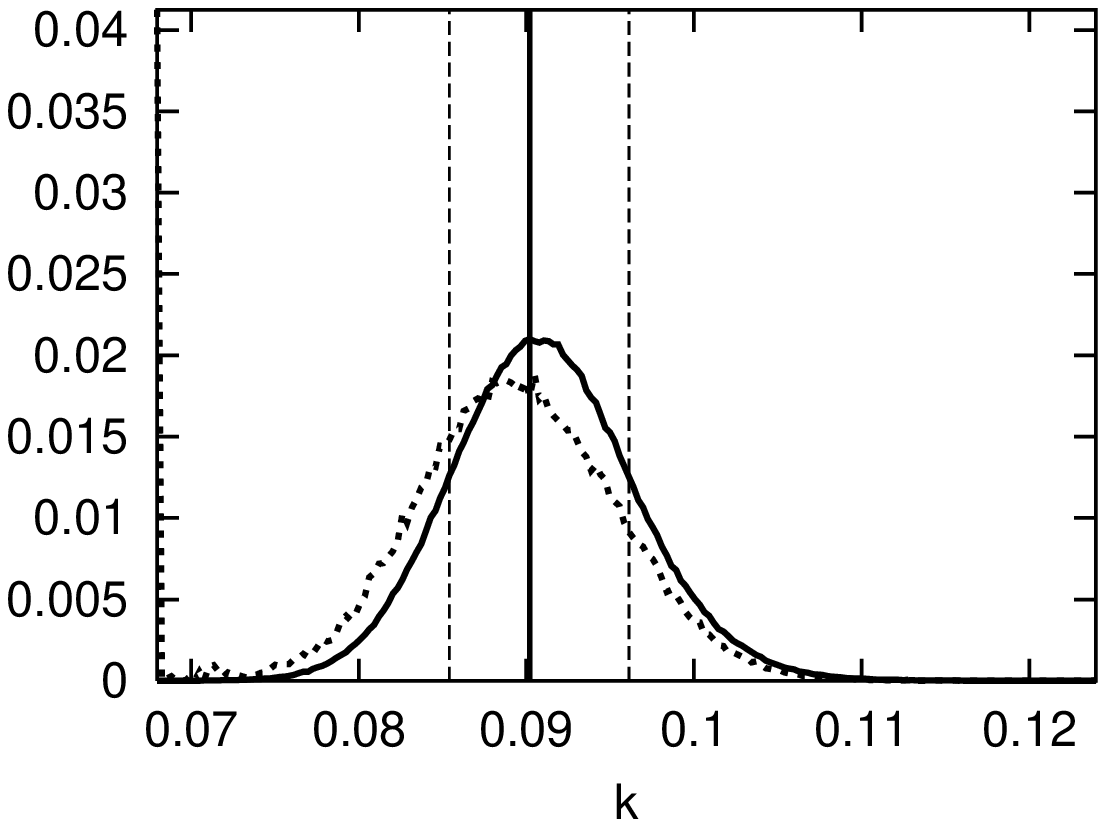}
\includegraphics[width=1.55in, height = 3.0in, angle=270]{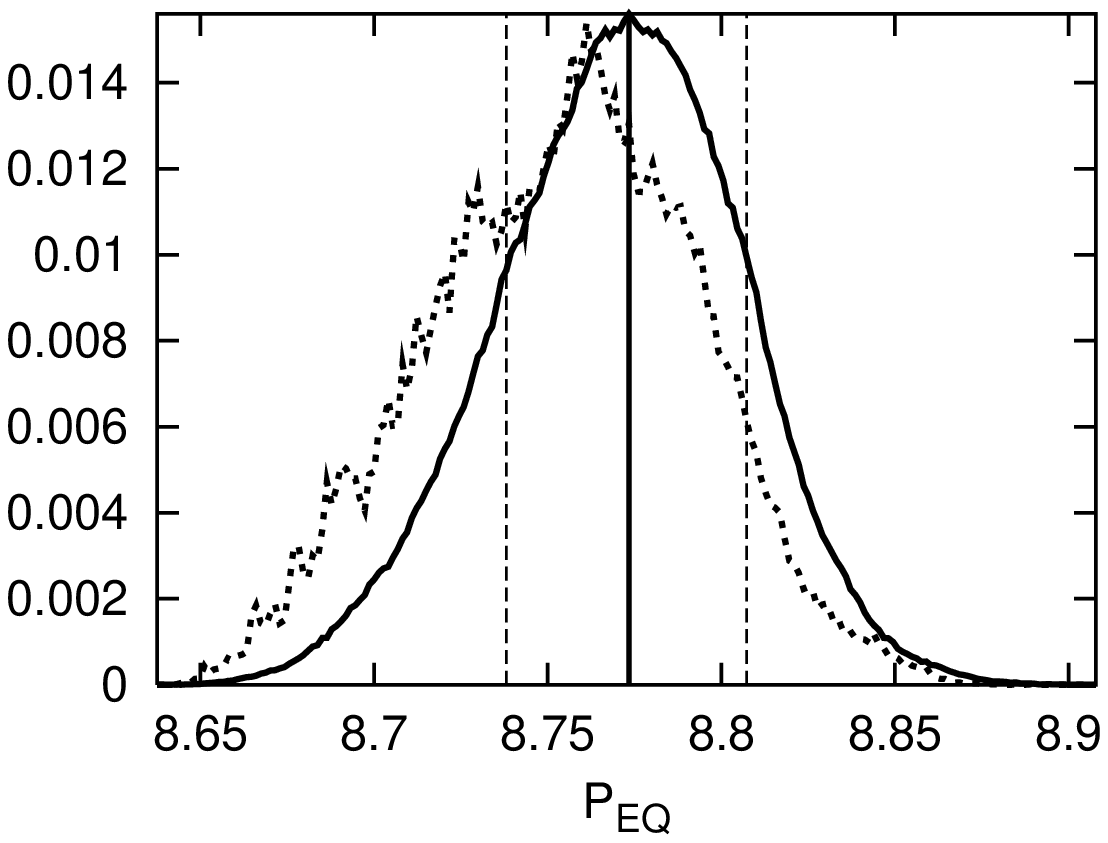}
\caption{
	{The marginalized likelihood for each of the fitted parameters is shown by the thick curve.
	The thin dotted curve shows the mean likelihood. The best-fitting value (the peak of the distribution) and the 68\% credible
	regions determined from the marginalized likelihood are shown by the solid vertical, and dashed vertical lines.
	The unusual spike in the mean likelihood curve for a low value of k in the k histogram 
	is likely due to low numbers
	of statistics, as indicated by the negligible value of the marginal likelihood curve.}  
\label{FigMCMC}
}
\end{figure}

\begin{figure}
\epsscale{1.0}

\includegraphics[width=1.85in, height = 3.5in, angle=270]{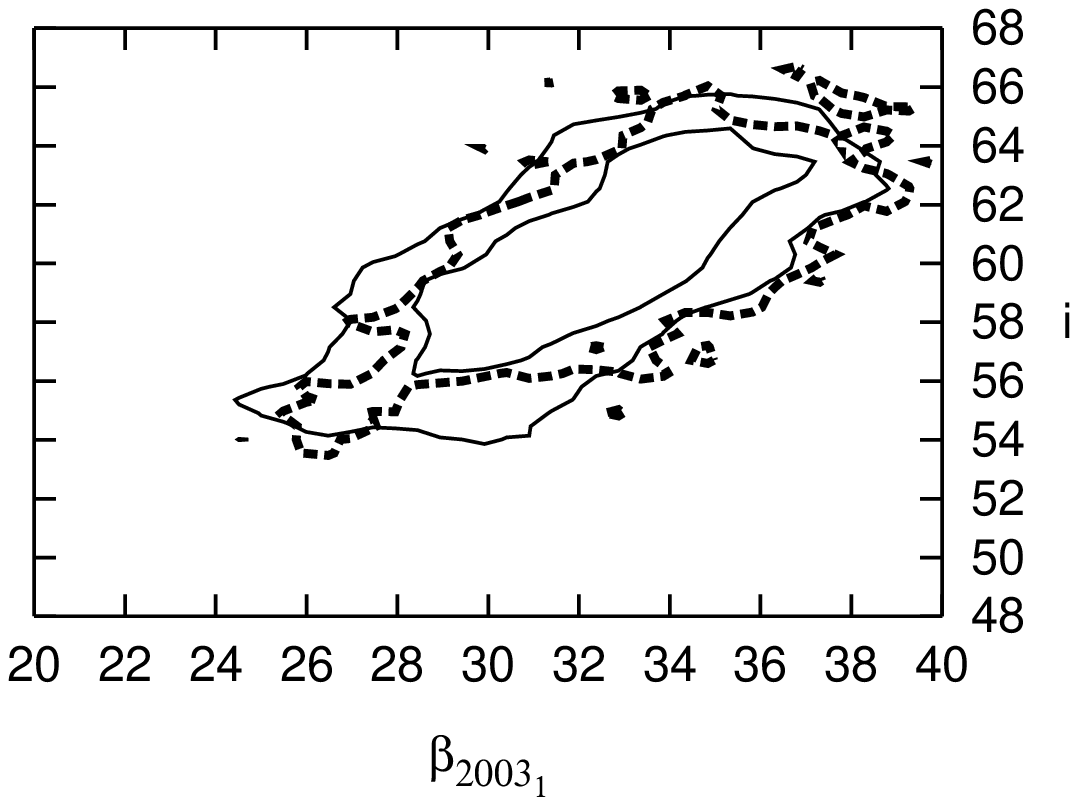}
\includegraphics[width=1.85in, height = 3.5in, angle=270]{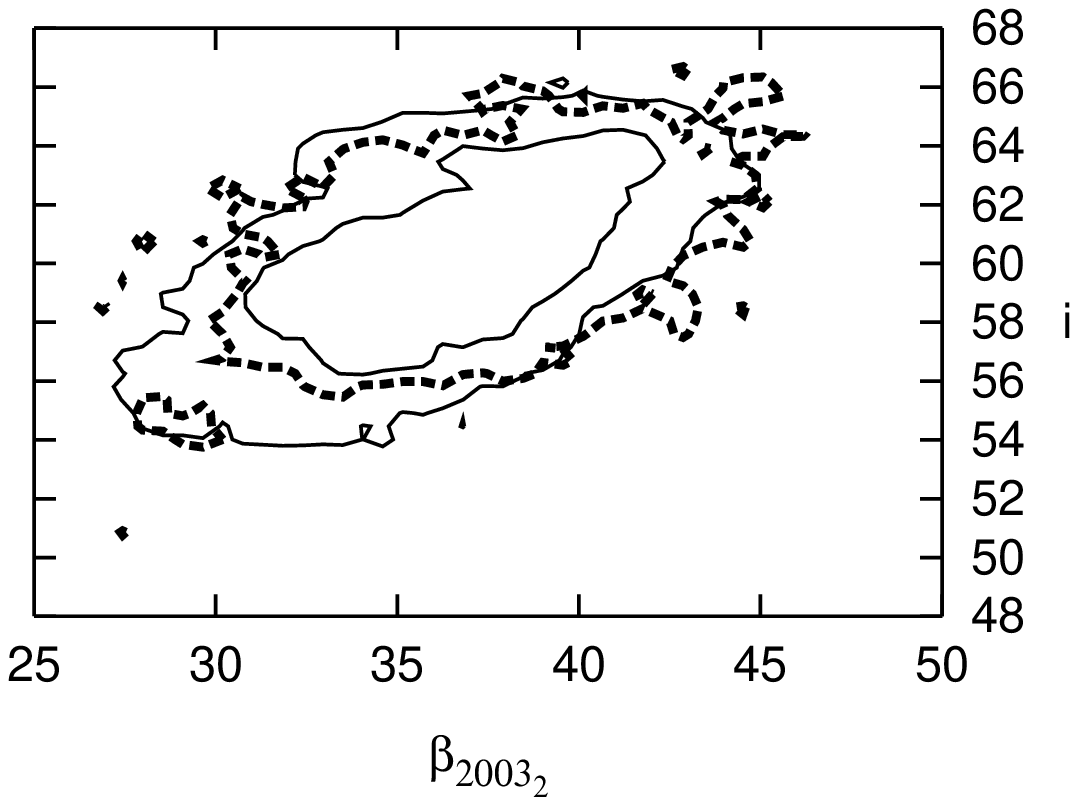}
\includegraphics[width=1.85in, height = 3.5in, angle=270]{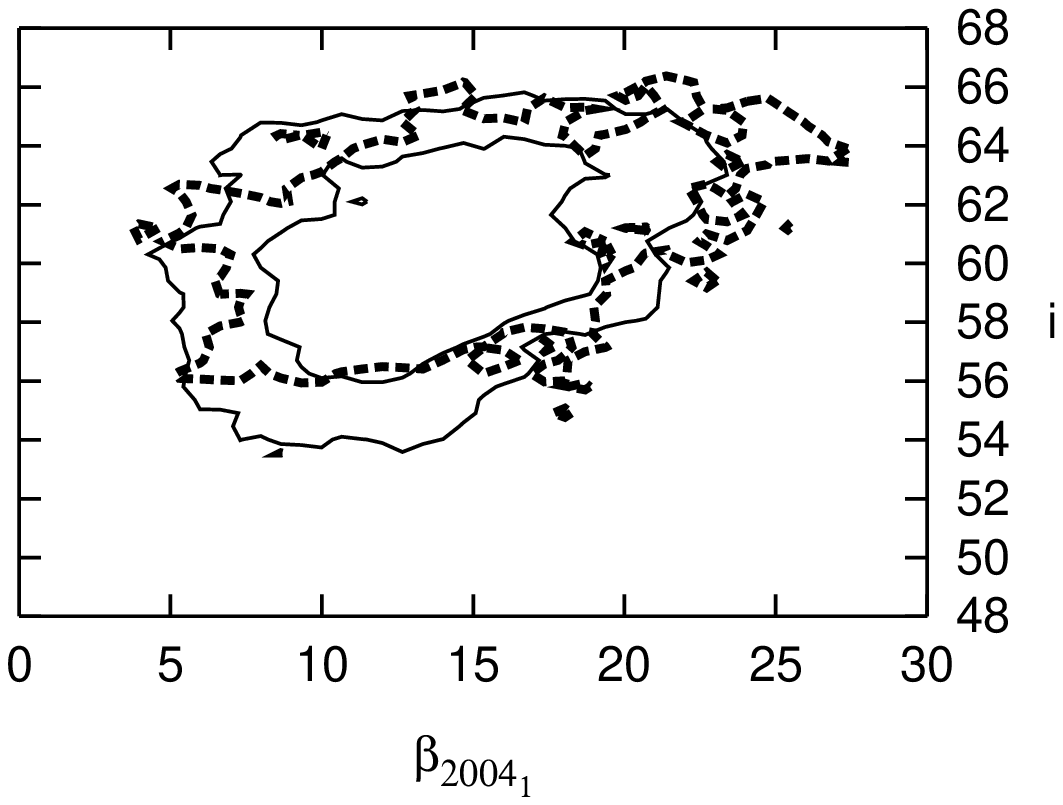}
\includegraphics[width=1.85in, height = 3.5in, angle=270]{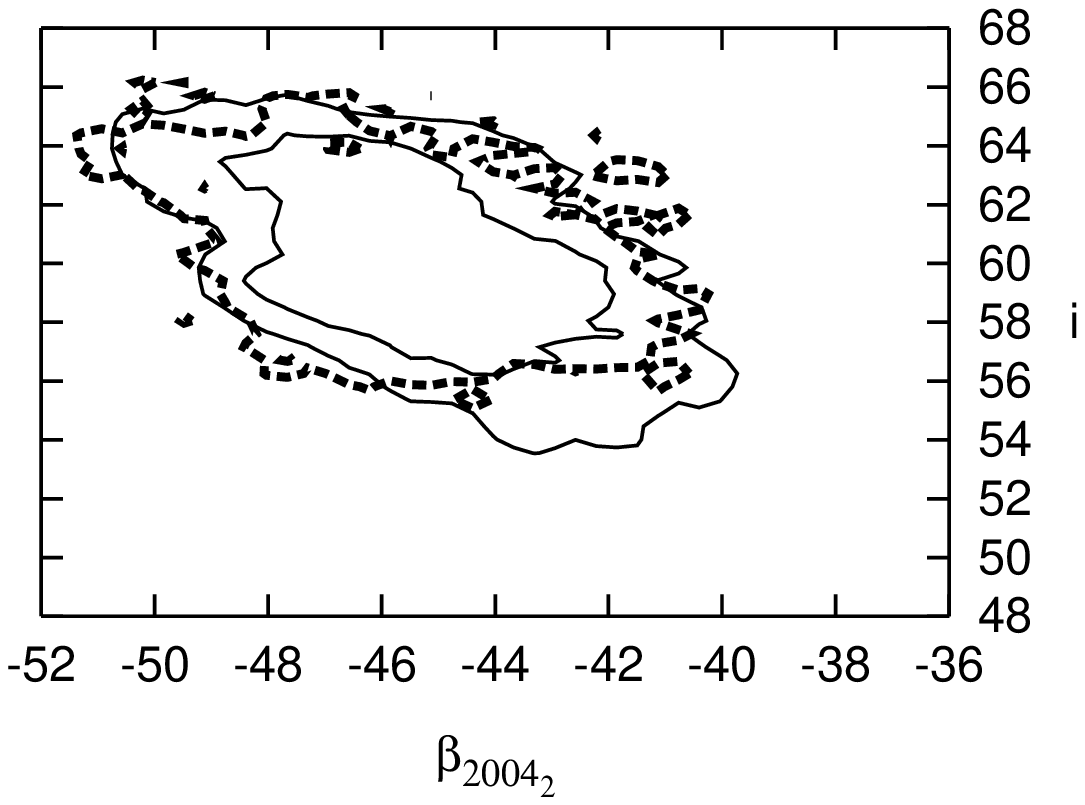}
\includegraphics[width=1.85in, height = 3.5in, angle=270]{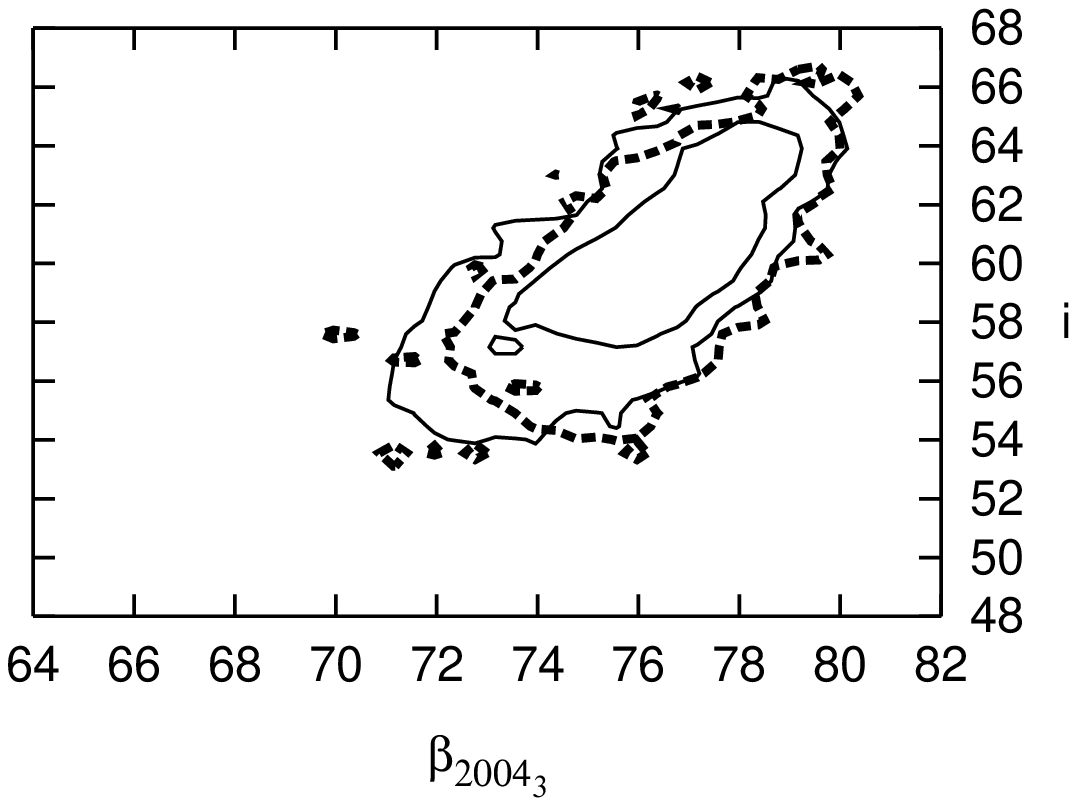}
\includegraphics[width=1.85in, height = 3.5in, angle=270]{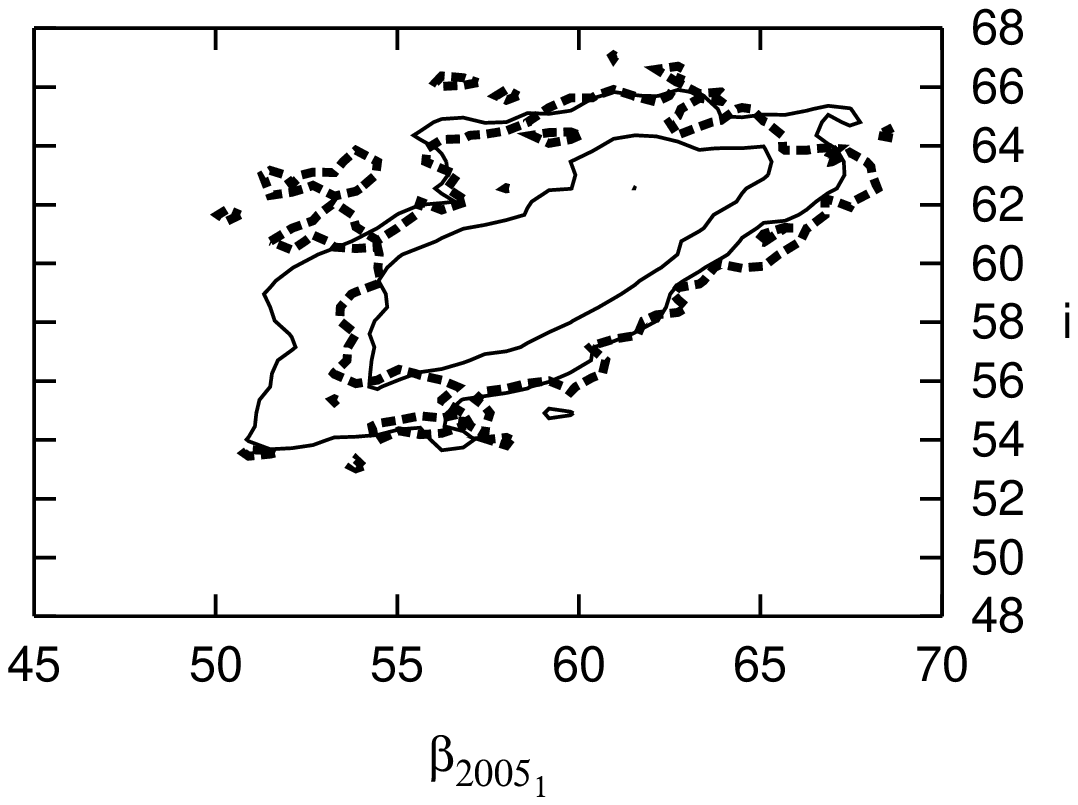}
\includegraphics[width=1.85in, height = 3.5in, angle=270]{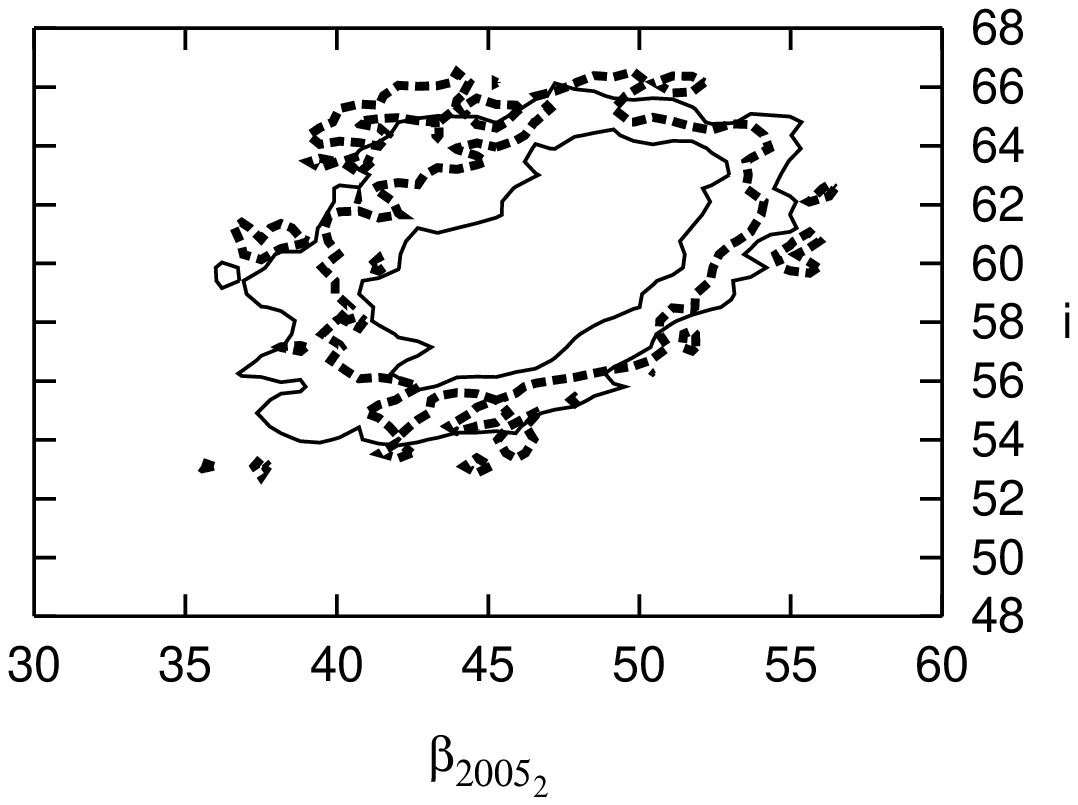}
\includegraphics[width=1.85in, height = 3.5in, angle=270]{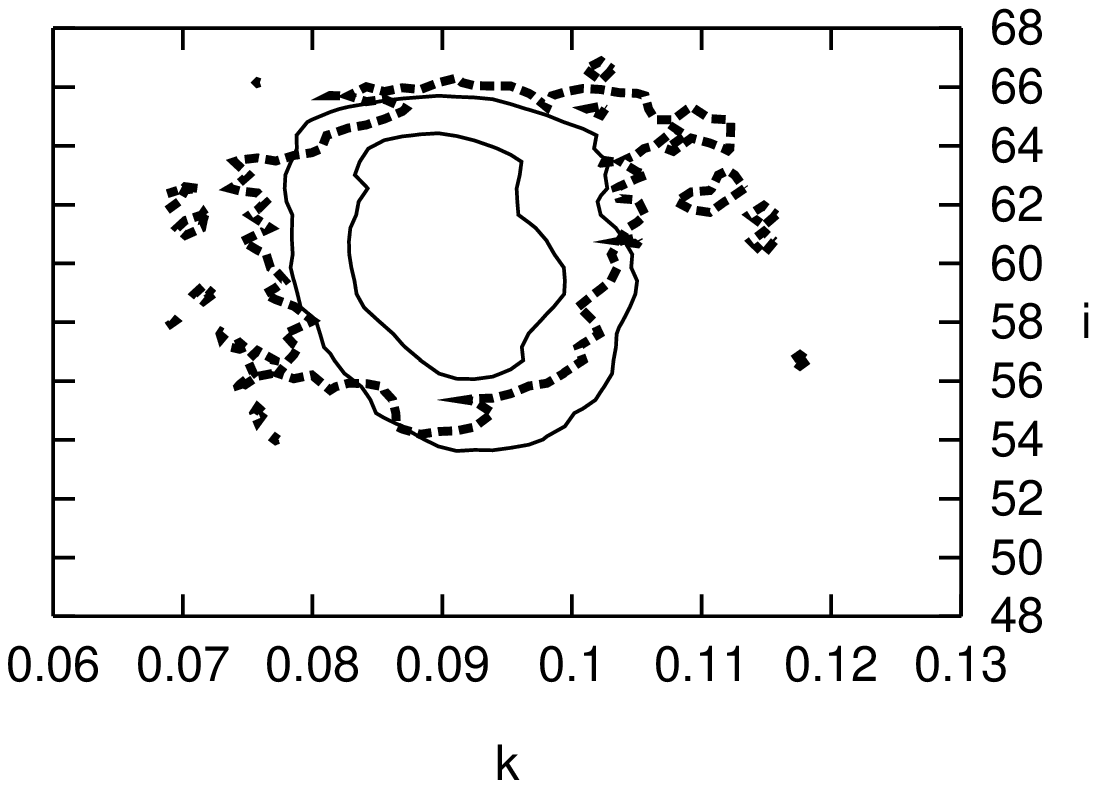}
\caption{
	{The solid lines are the 68\% and 95\% credible regions of the marginalized likelihood for various parameters. 
	The 68\% mean likelihood credible regions are shown by dotted contours (see text).
	Obviously there is little correlation between $k$ and $i$ in the bottom right panel. }
\label{FigDegen}
}
\end{figure}

\begin{figure}
\epsscale{1.0}
\begin{center}
\includegraphics[width=1.55in, height = 1.55in]{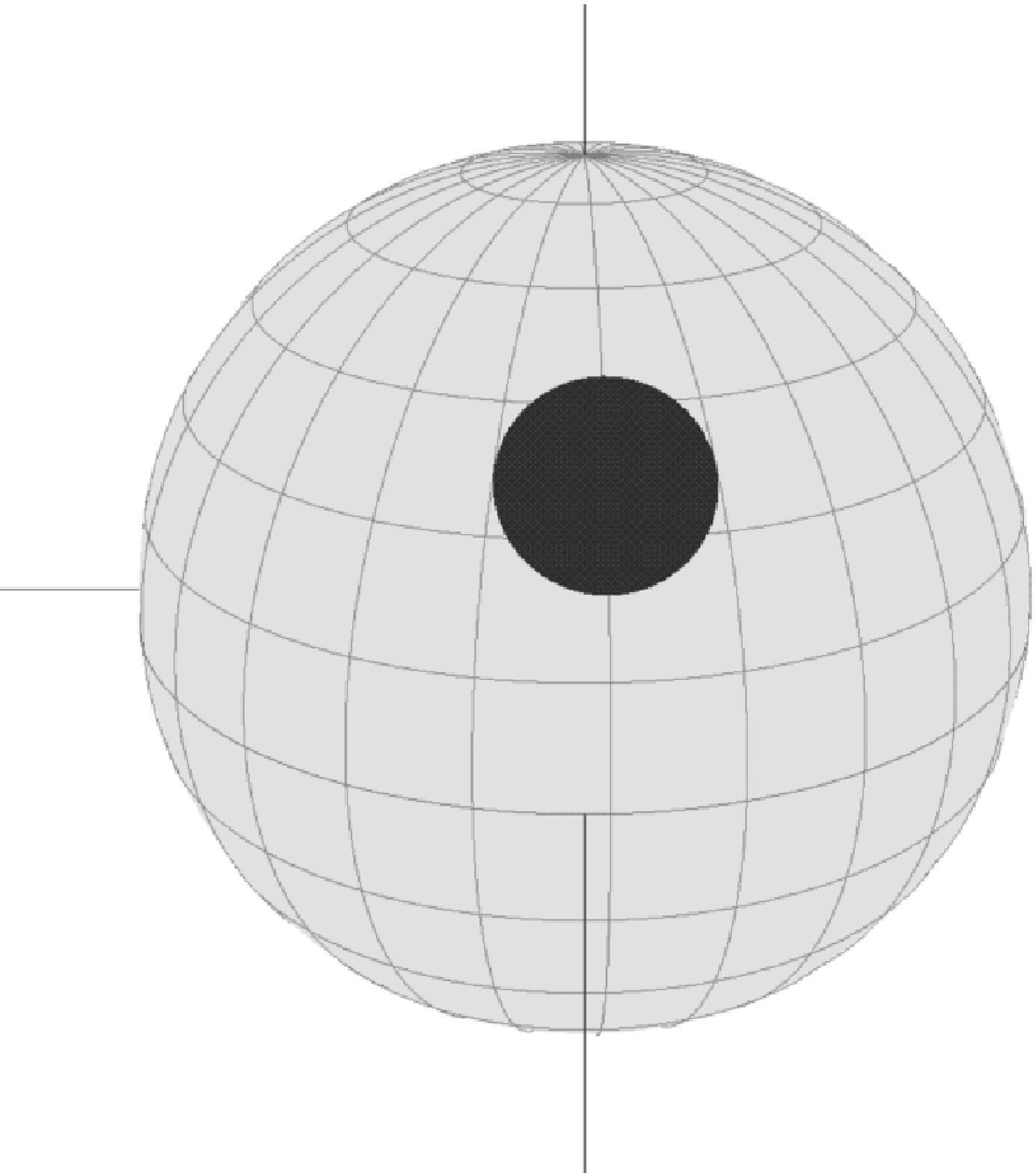}
\includegraphics[width=1.55in, height = 1.55in]{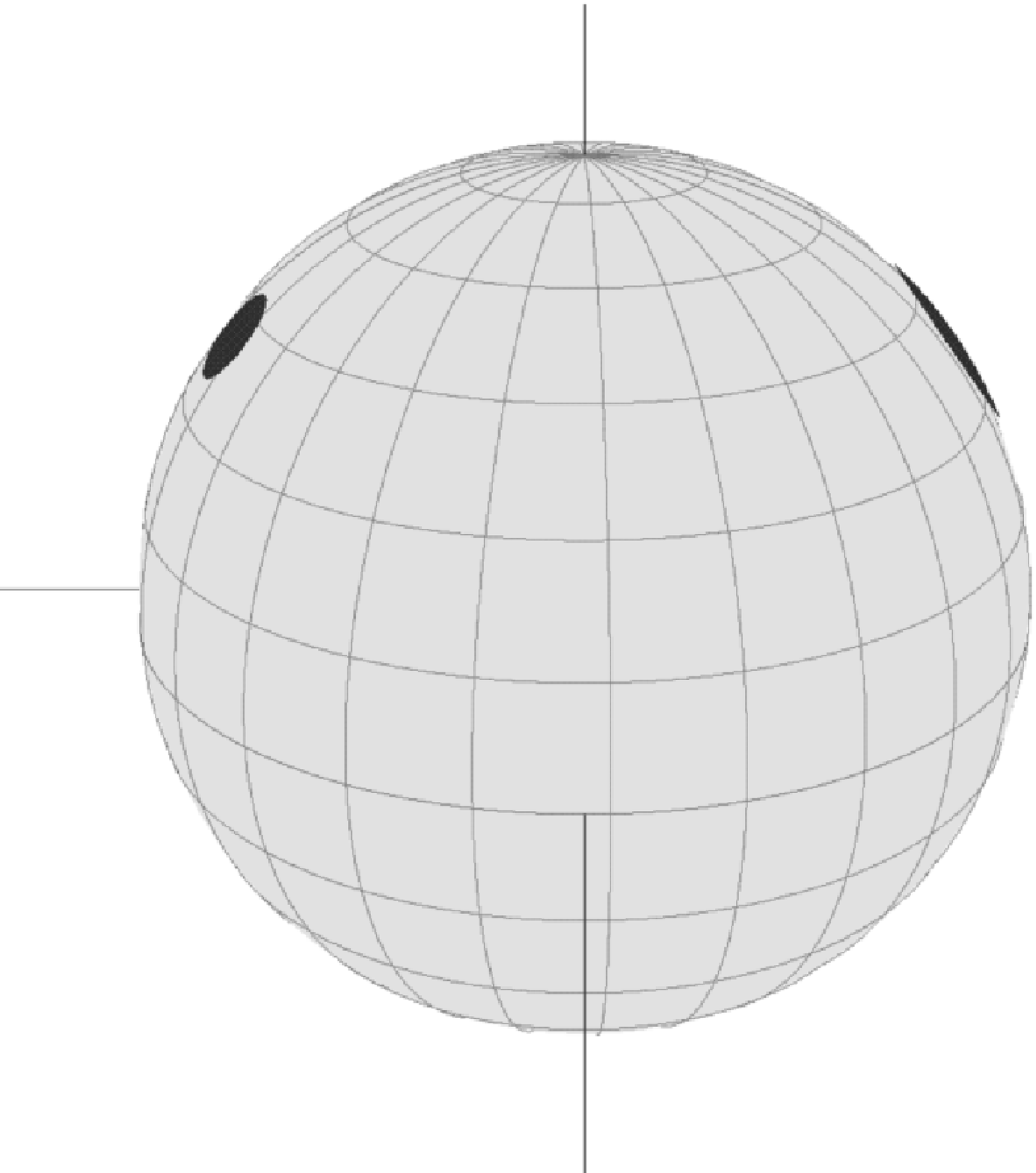}
\includegraphics[width=1.55in, height = 1.55in]{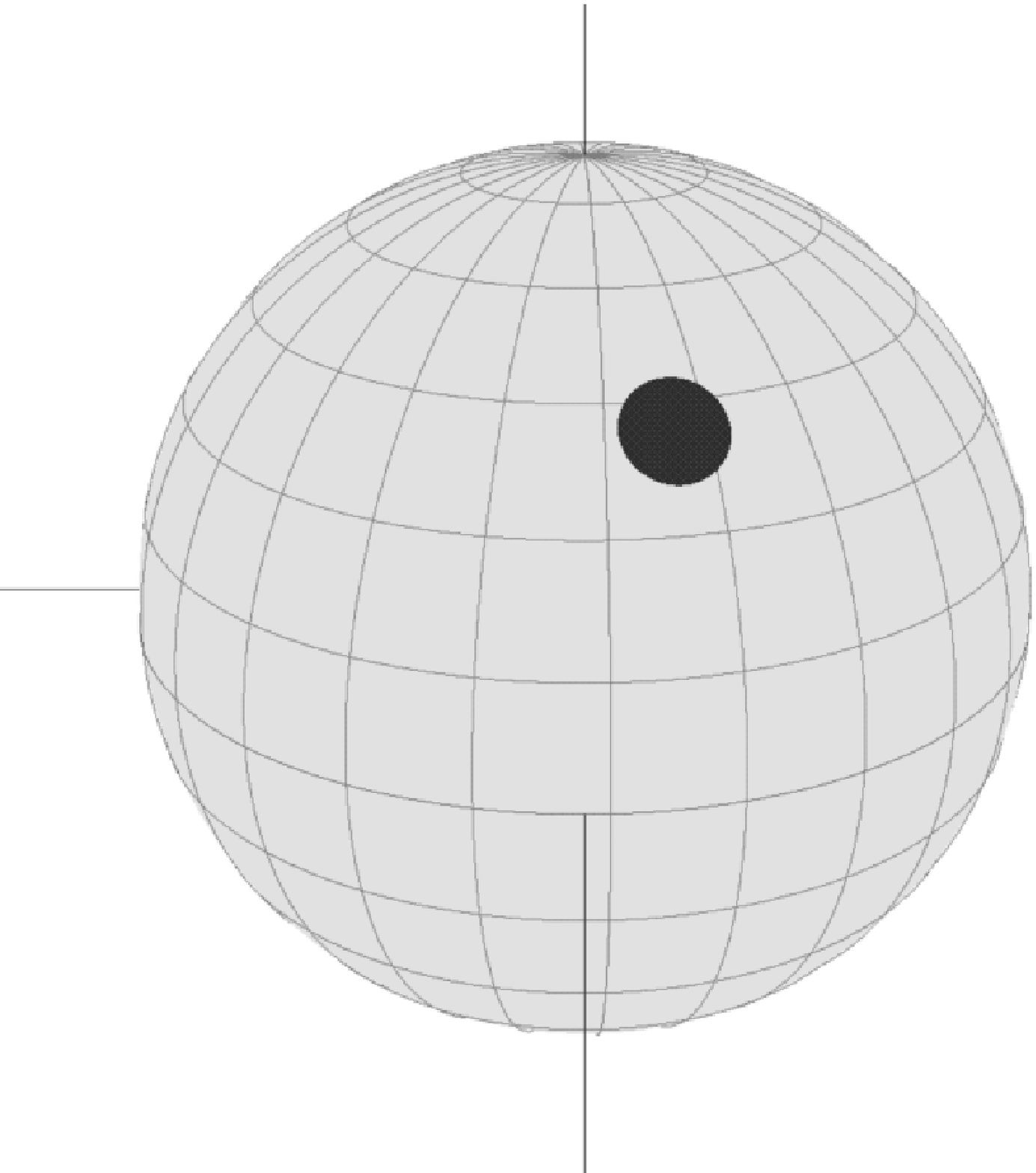}
\includegraphics[width=1.55in, height = 1.55in]{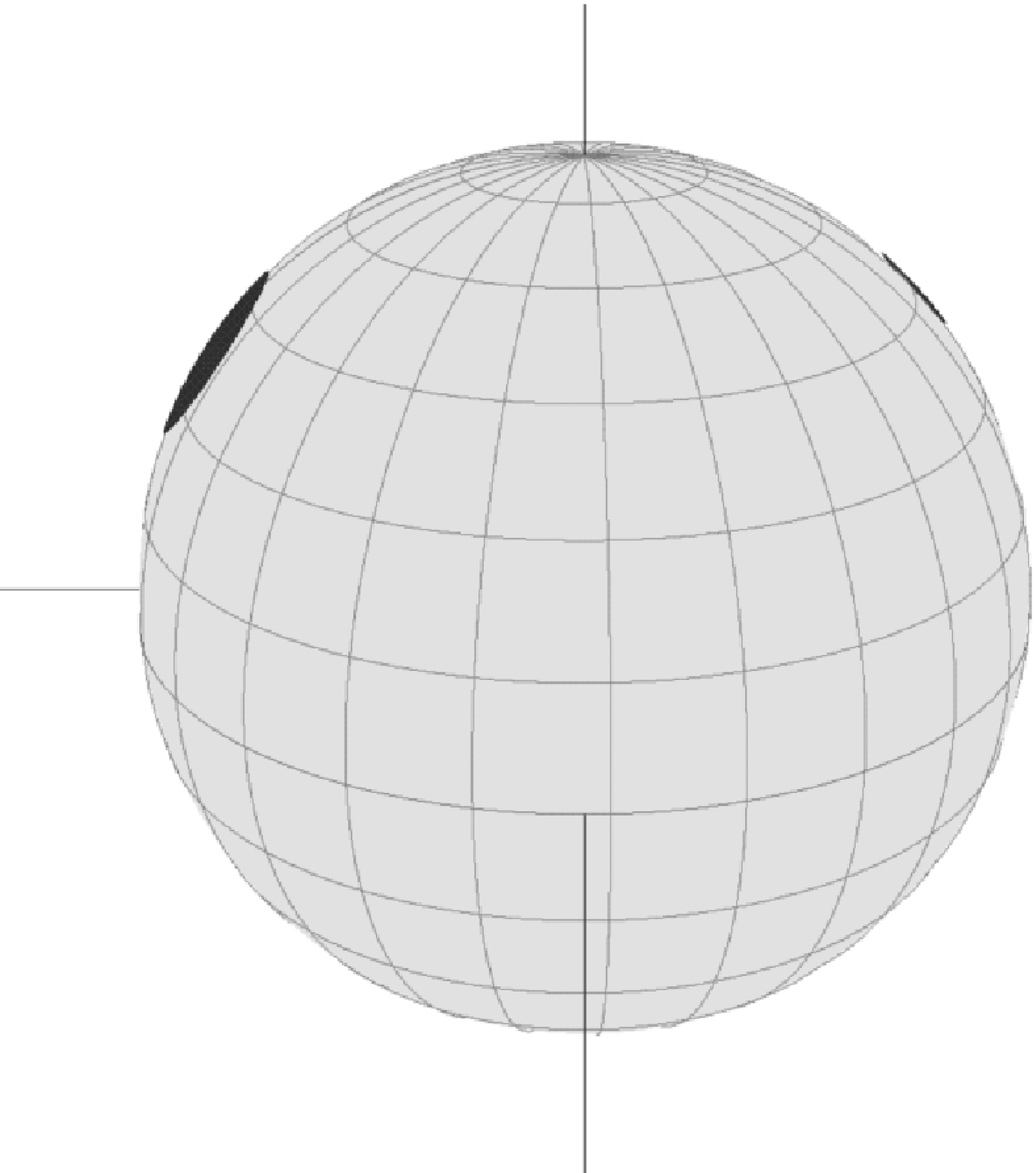}
\end{center}
\includegraphics[width=6.0in, height = 7.0in, angle=270]{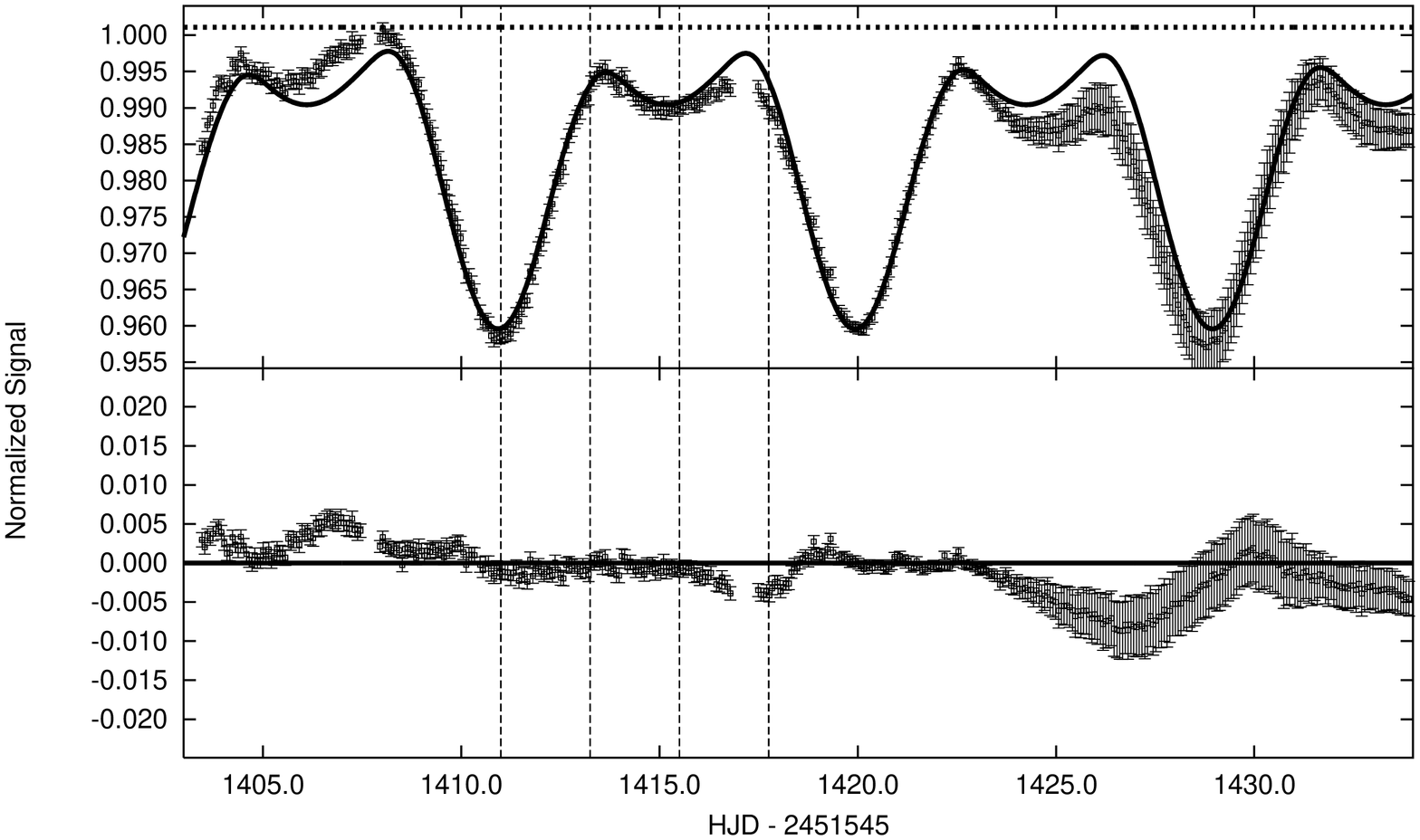}
\caption{
	{The best-fitting two spot solution for $\kappa^1$ Ceti in 2003 (rotating counter-clockwise from top) and seen from the line of sight at phases 0.00, 0.25, 0.50, and 0.75  (from left) of the first spot. 
	Middle: the {\it MOST} light curve with errors (see text). The continuous line is the solution from the ``Minimum $\chi^{2}$'' column of Table \ref{TableMCMC}. The dotted line indicates the unspotted normalized signal of the star ($U$=\UOThreemin). Vertical dashed lines indicate phases 0.00, 0.25, 0.50 and 0.75.
	Bottom: residuals from the model on the same scale.}
\label{Fig2003}
}
\end{figure}

\begin{figure}
\epsscale{1.0}
\begin{center}

\includegraphics[width=1.55in, height = 1.55in]{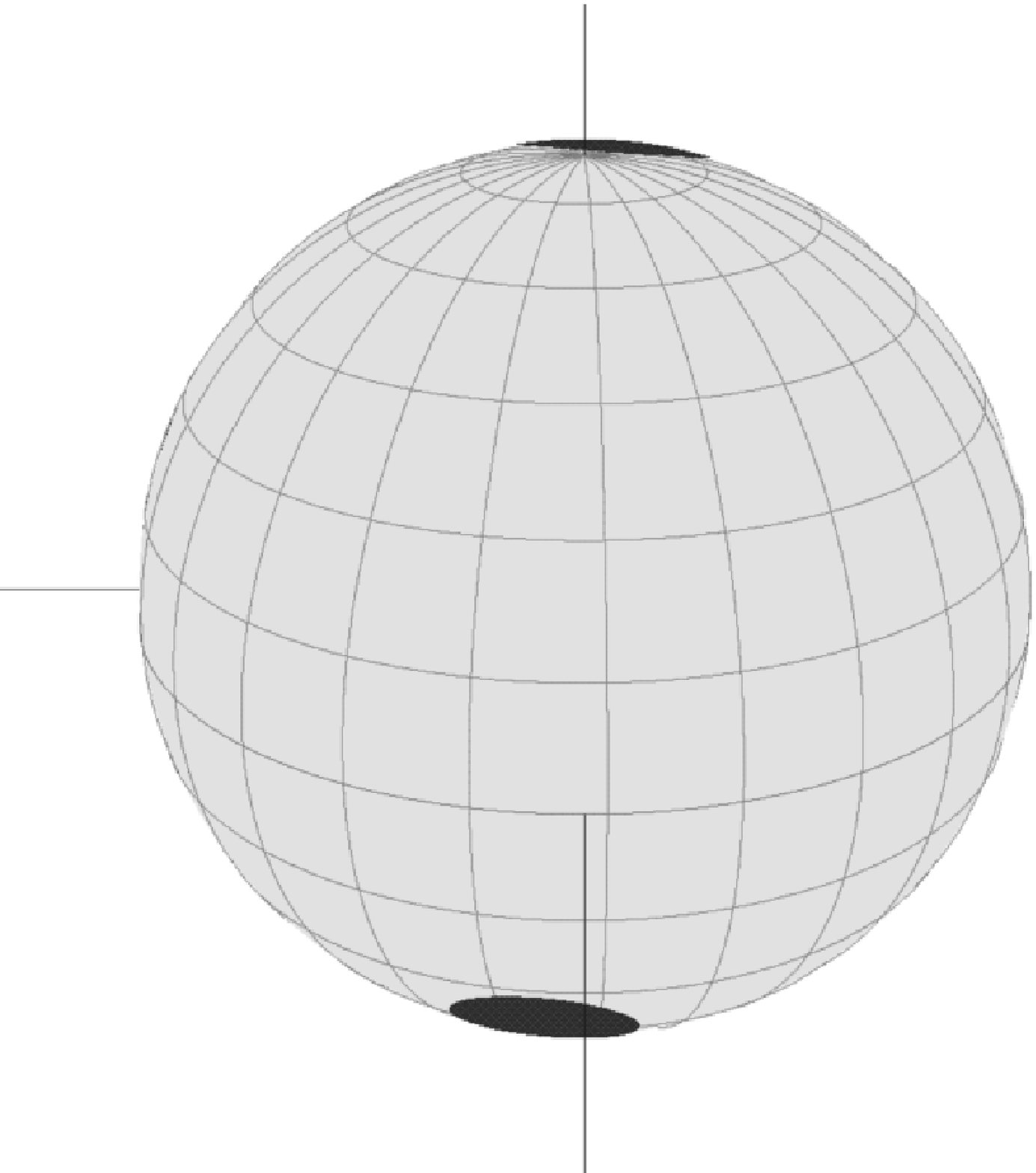}
\includegraphics[width=1.55in, height = 1.55in]{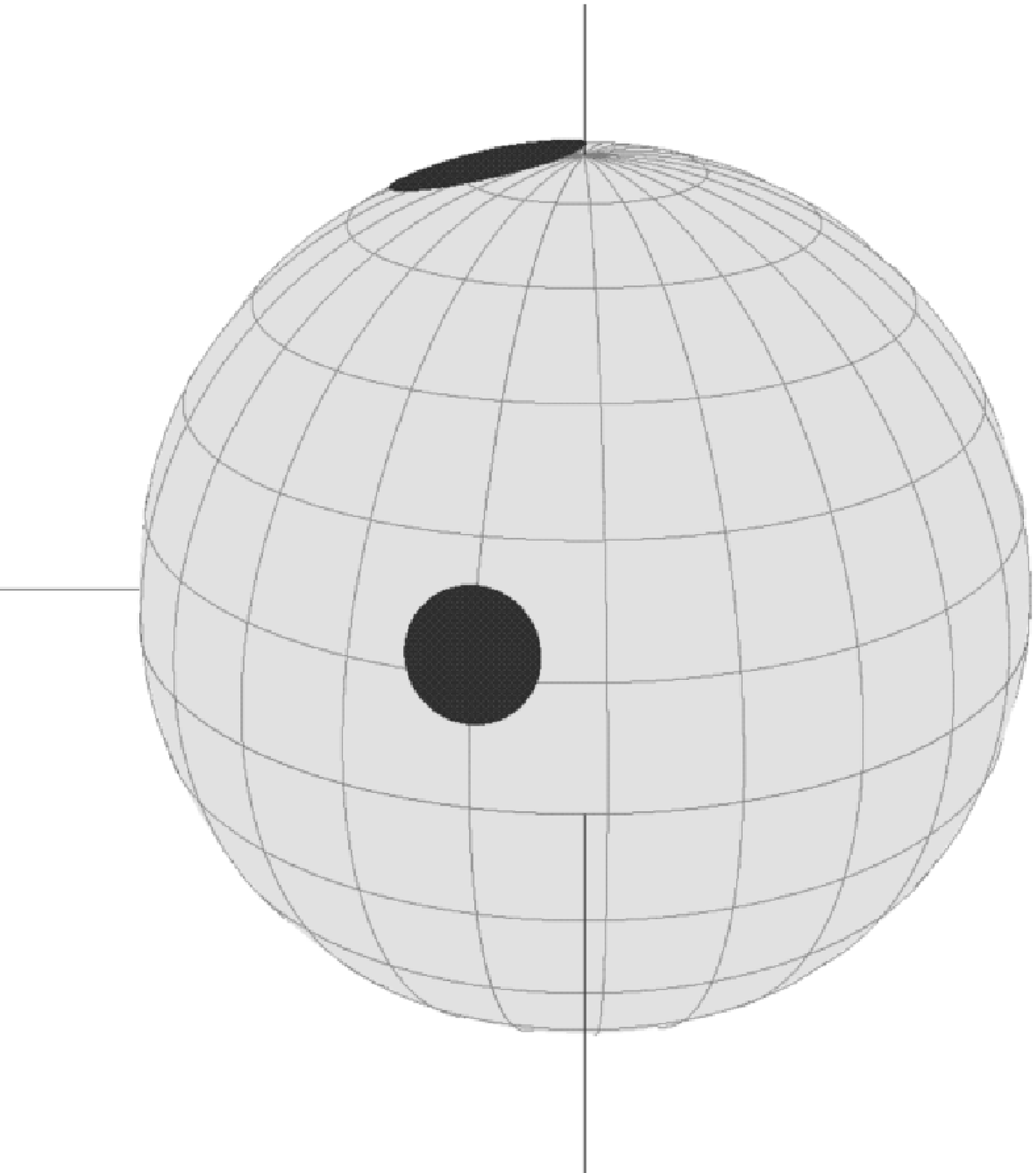}
\includegraphics[width=1.55in, height = 1.55in]{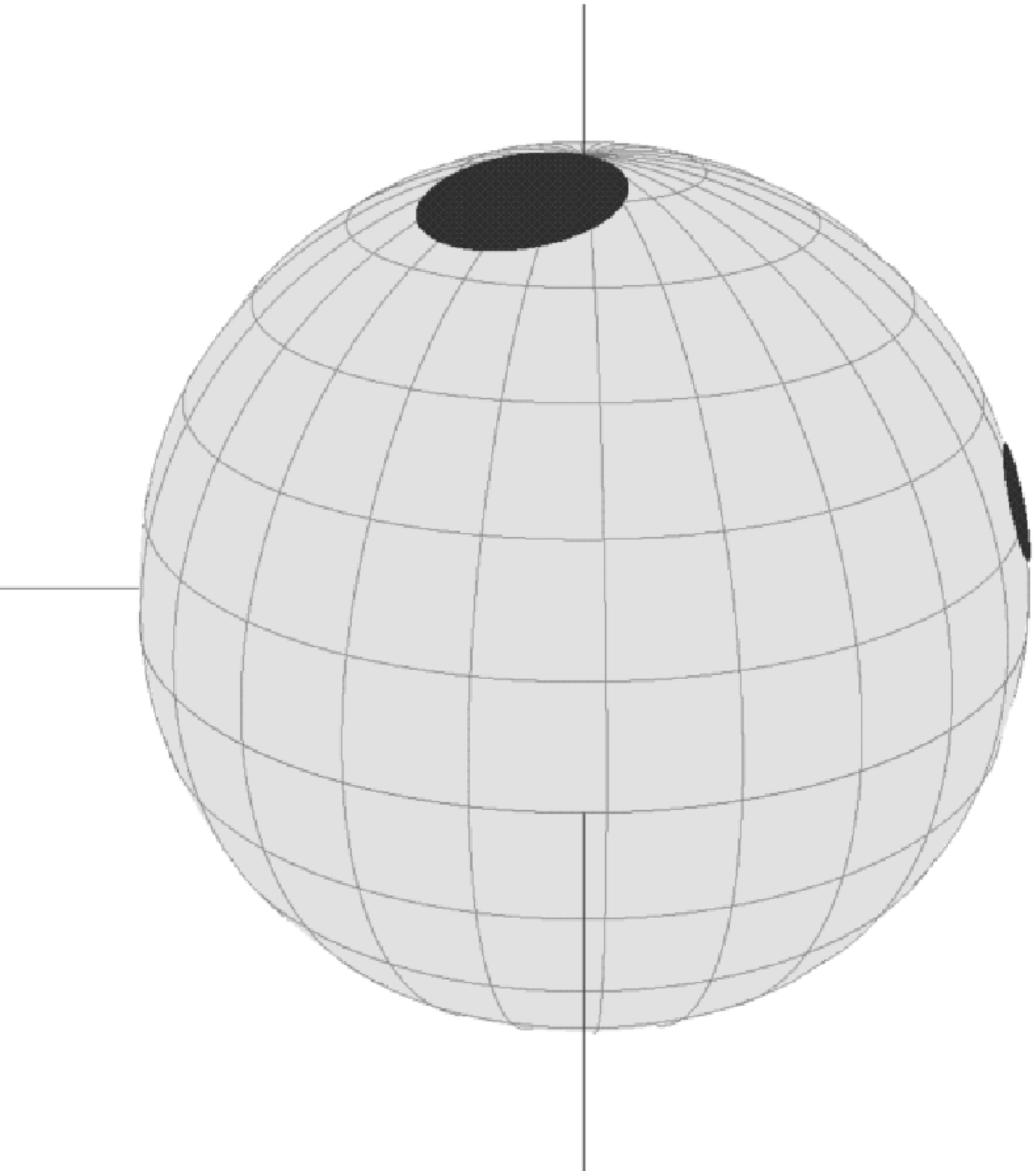}
\includegraphics[width=1.55in, height = 1.55in]{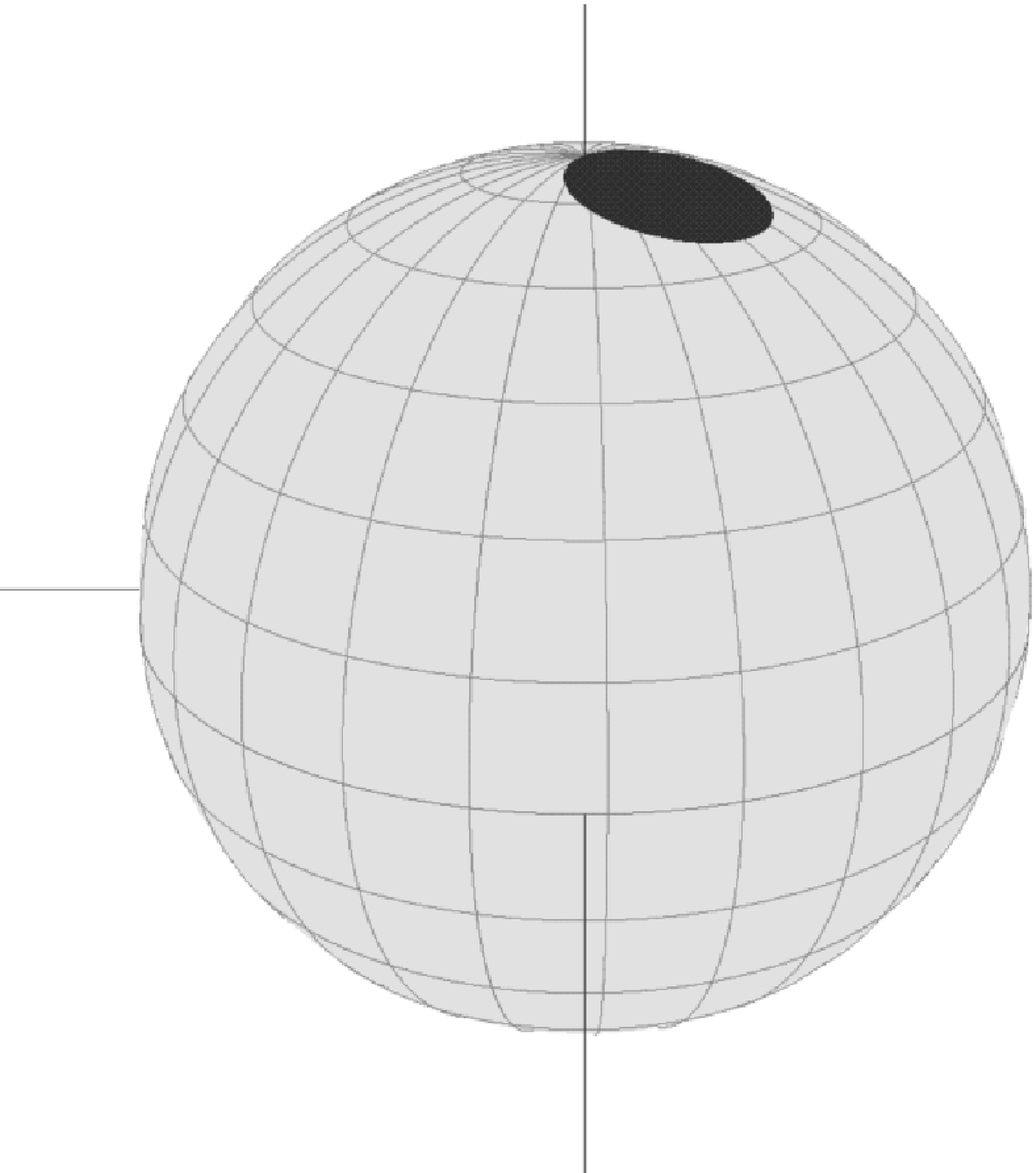}
\end{center}
\includegraphics[width=6.0in, height = 7.0in, angle=270]{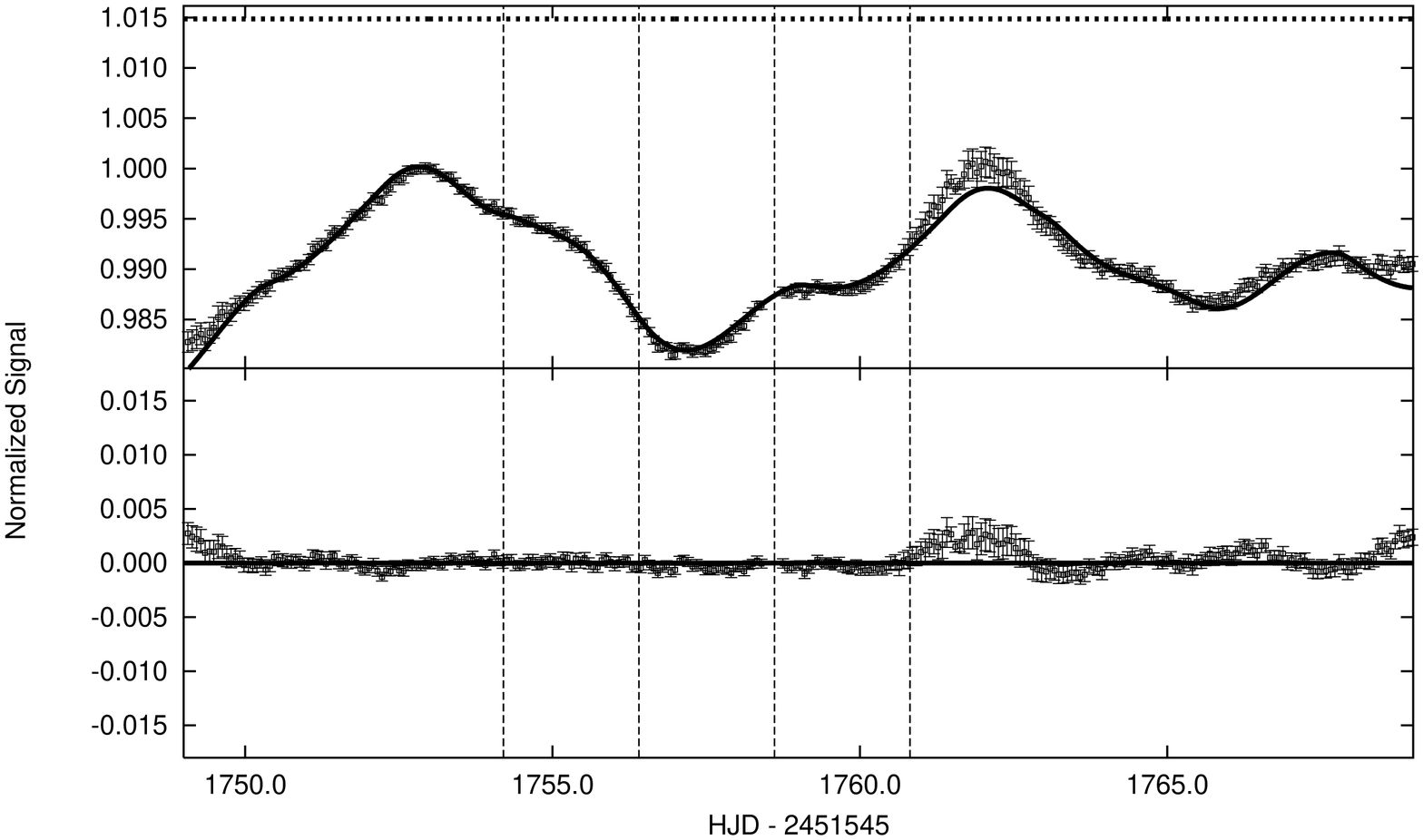}
\caption{
	{The best-fitting three spot solution for $\kappa^1$  Ceti in 2004 (rotating counter-clockwise from top) and seen from the line of sight at phases 0.00, 0.25, 0.50, and 0.75  (from left) of the first spot. Middle: the {\it MOST} light curve with errors (see text). The continuous line is the solution from the ``Minimum $\chi^{2}$'' column of Table \ref{TableMCMC}. The dotted line indicates the unspotted normalized signal of the star ($U$=\UOFourmin). Vertical dashed lines indicate phases 0.00, 0.25, 0.50 and 0.75. Bottom: residuals from the model on the same scale.} 
\label{Fig2004}
}
\end{figure}

\begin{figure}
\epsscale{1.0}
\begin{center}

\includegraphics[width=1.55in, height = 1.55in]{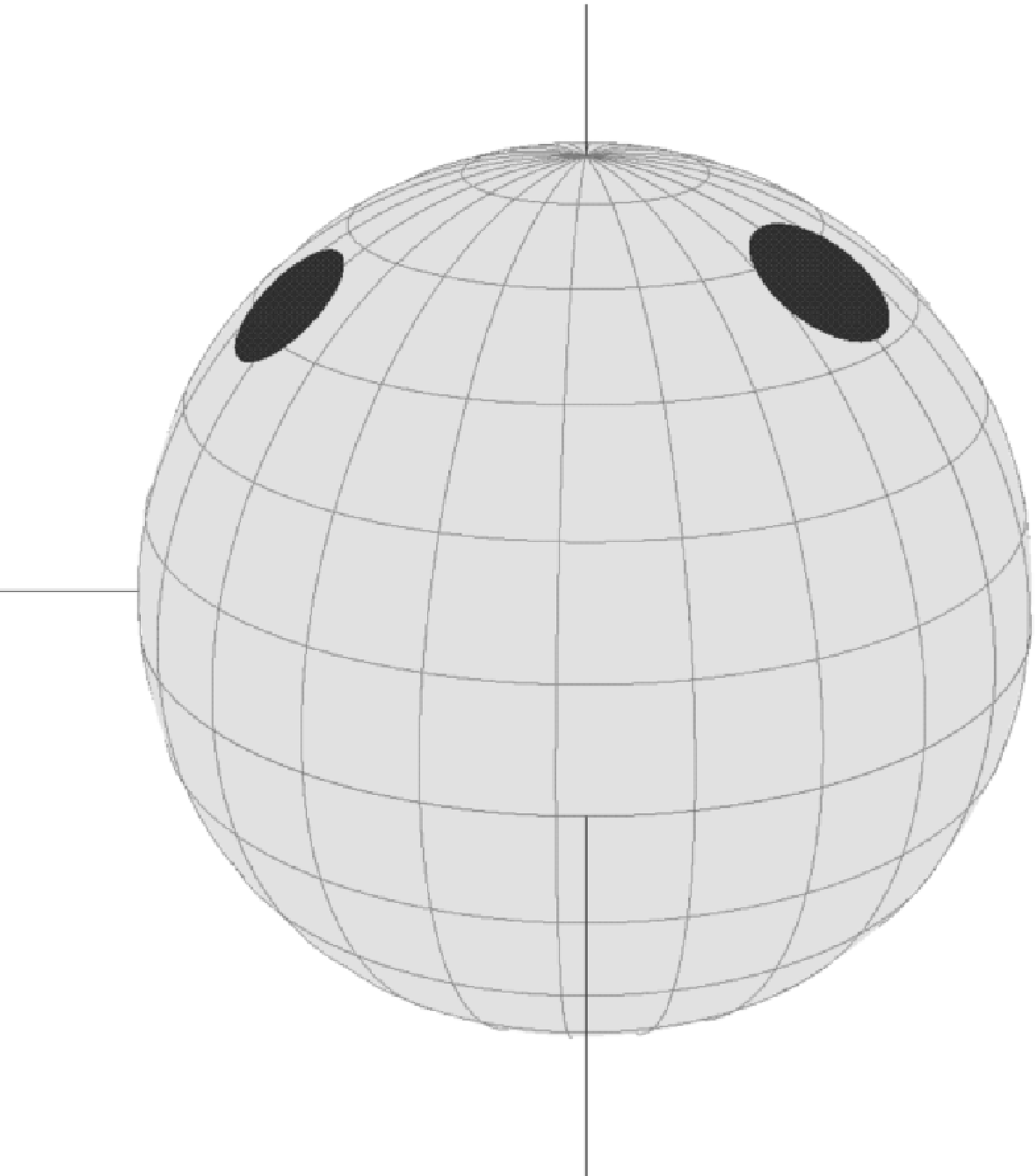}
\includegraphics[width=1.55in, height = 1.55in]{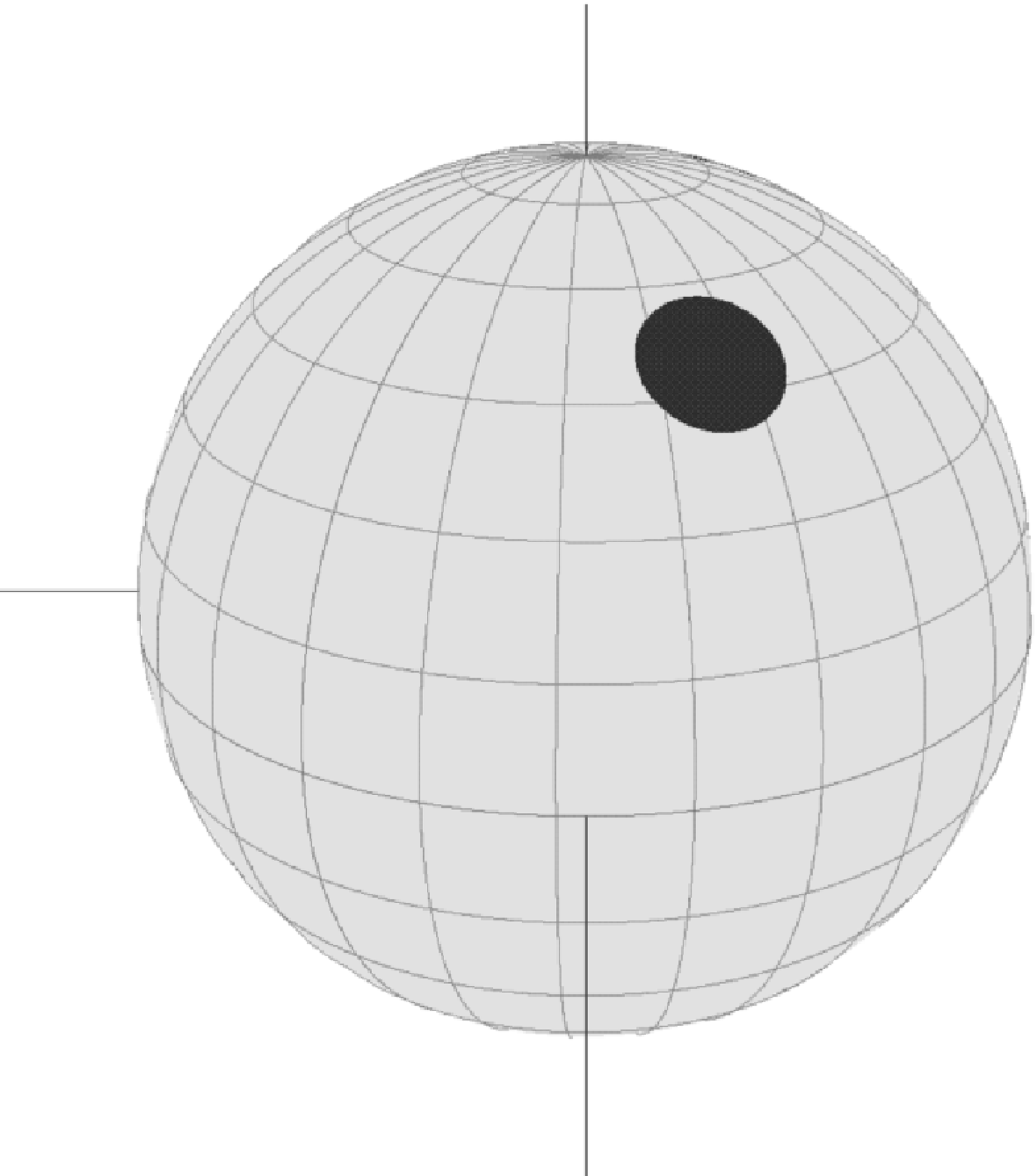}
\includegraphics[width=1.55in, height = 1.55in]{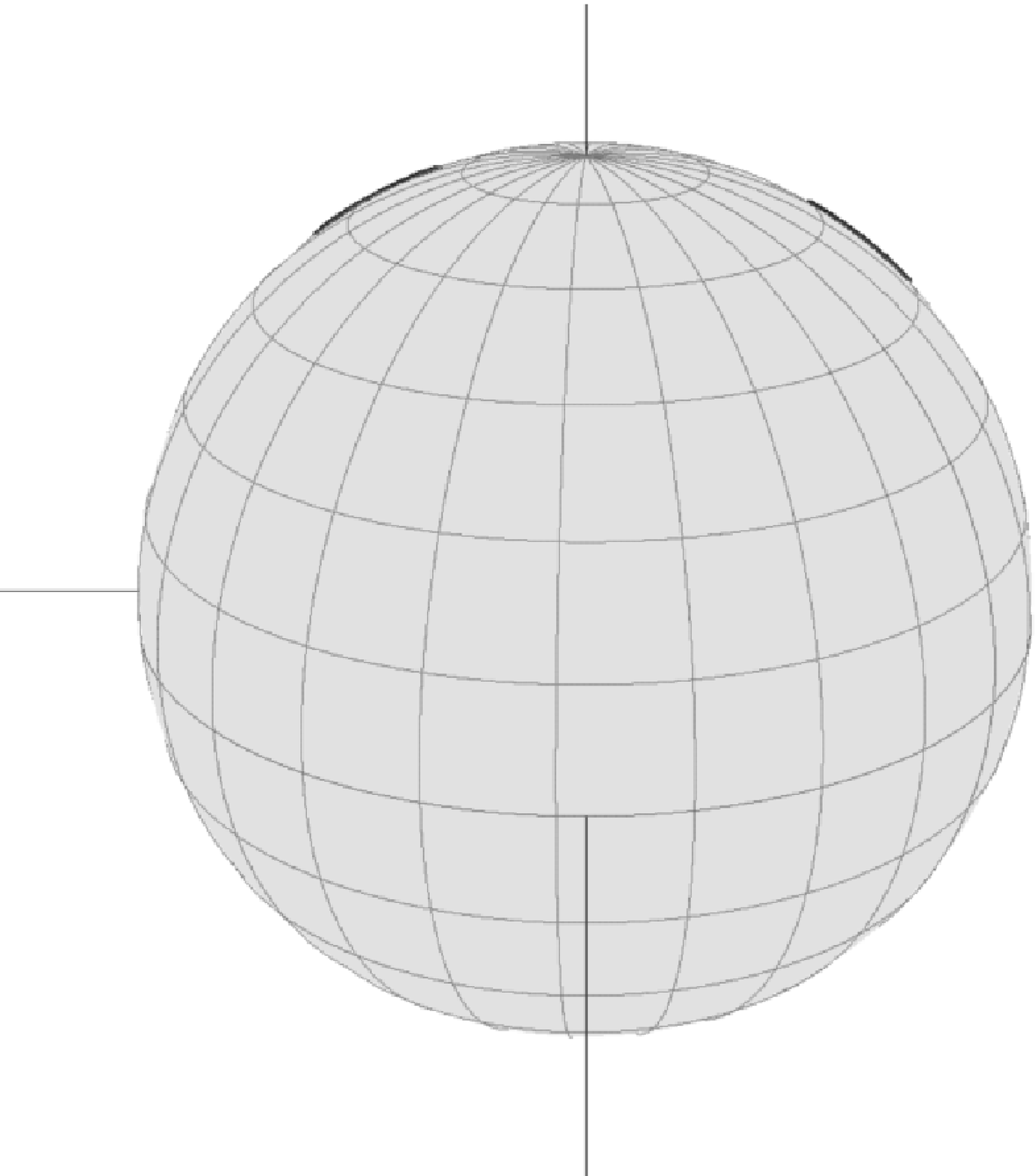}
\includegraphics[width=1.55in, height = 1.55in]{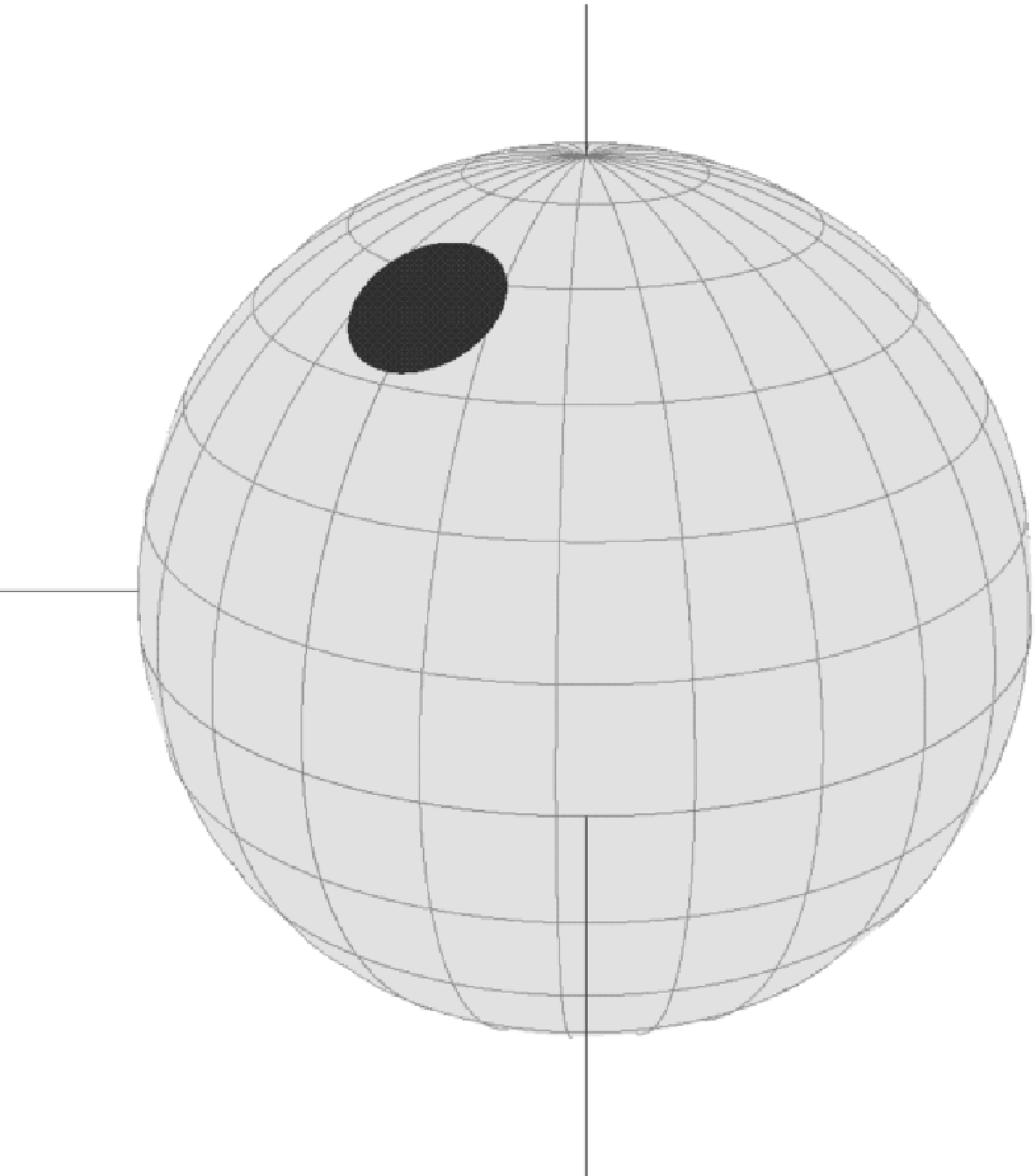}
\end{center}
\includegraphics[width=6.0in, height = 7.0in, angle=270]{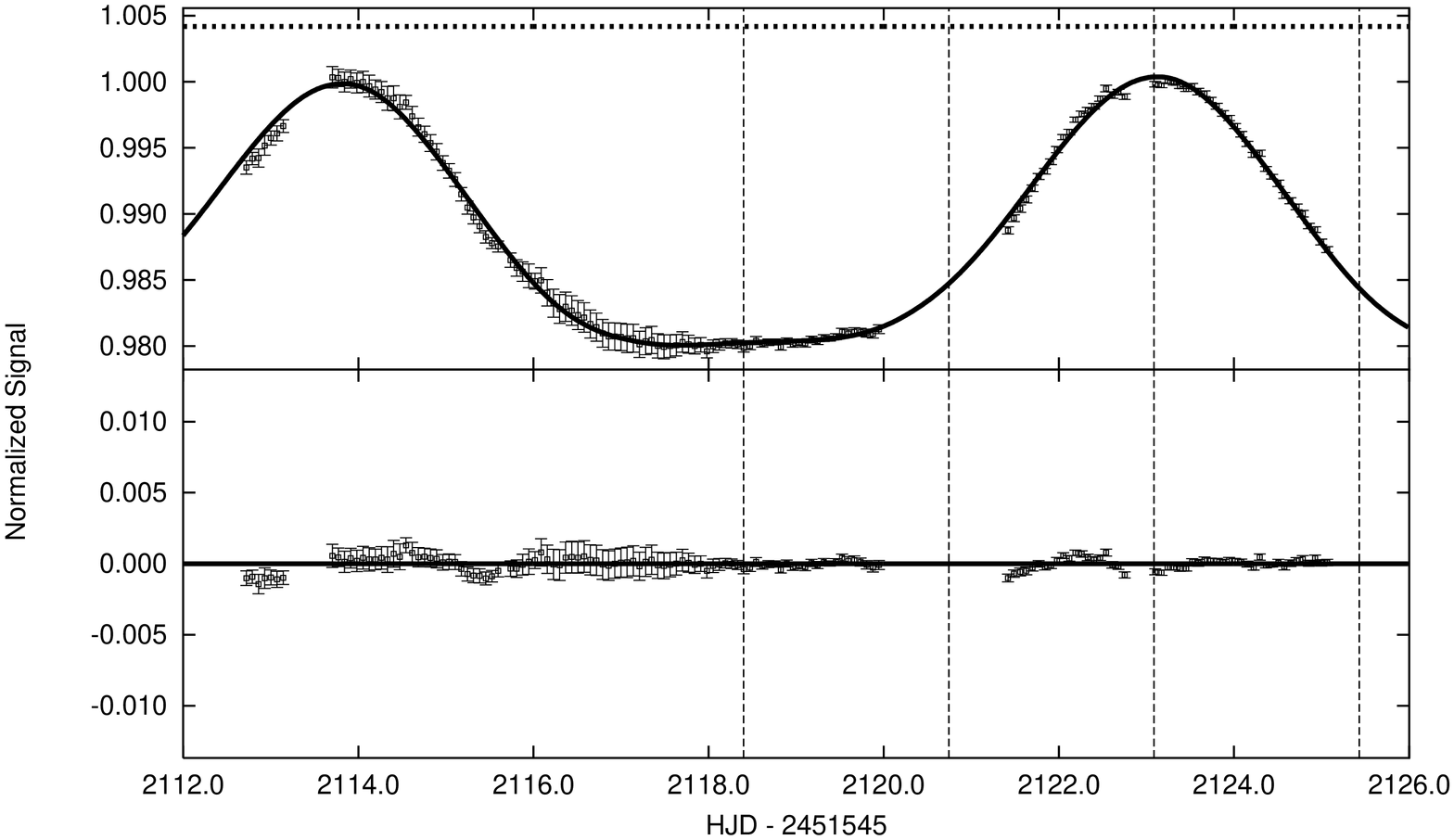}
\caption{
	{The best-fitting two spot solution for $\kappa^1$ in Ceti 2005 (rotating counter-clockwise from top) and seen from the line of sight at phases 0.00, 0.25, 0.50, and 0.75  (from left) of the first spot. 
	Middle: the {\it MOST} light curve with errors (see text). The continuous line is the solution from the ``Minimum $\chi^{2}$'' column of Table \ref{TableMCMC}. The dotted line indicates the unspotted normalized signal of the star ($U$=\UOFivemin). Vertical dashed lines indicate  phases 0.00, 0.25, 0.50 and 0.75.
	Bottom: residuals from the model on the same scale.}  
\label{Fig2005}
}
\end{figure}

\begin{figure}
\epsscale{1.0}
\includegraphics[width=3.50in, height = 6.5in, angle=270]{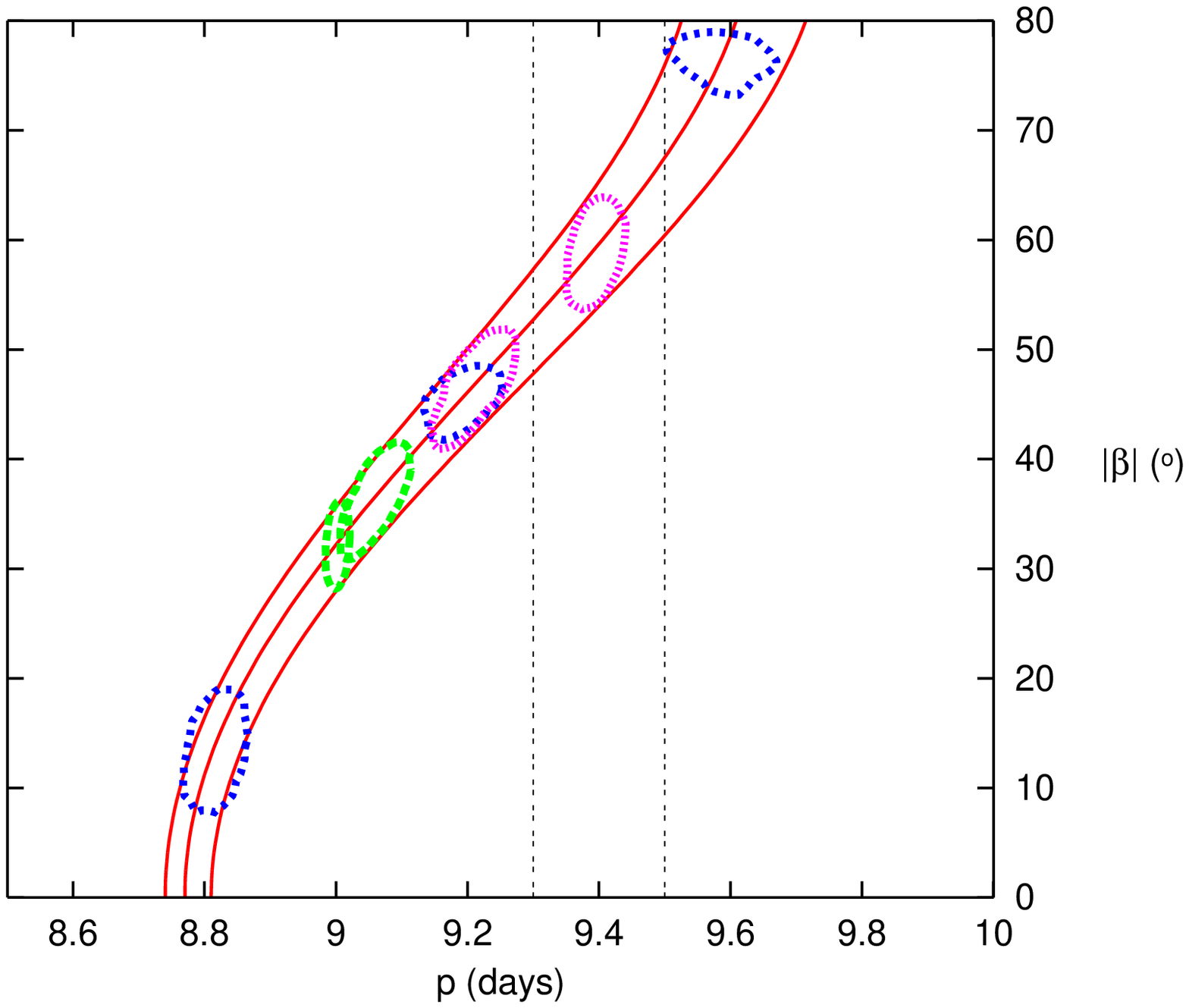}
\includegraphics[width=3.50in, height = 6.5in, angle=270]{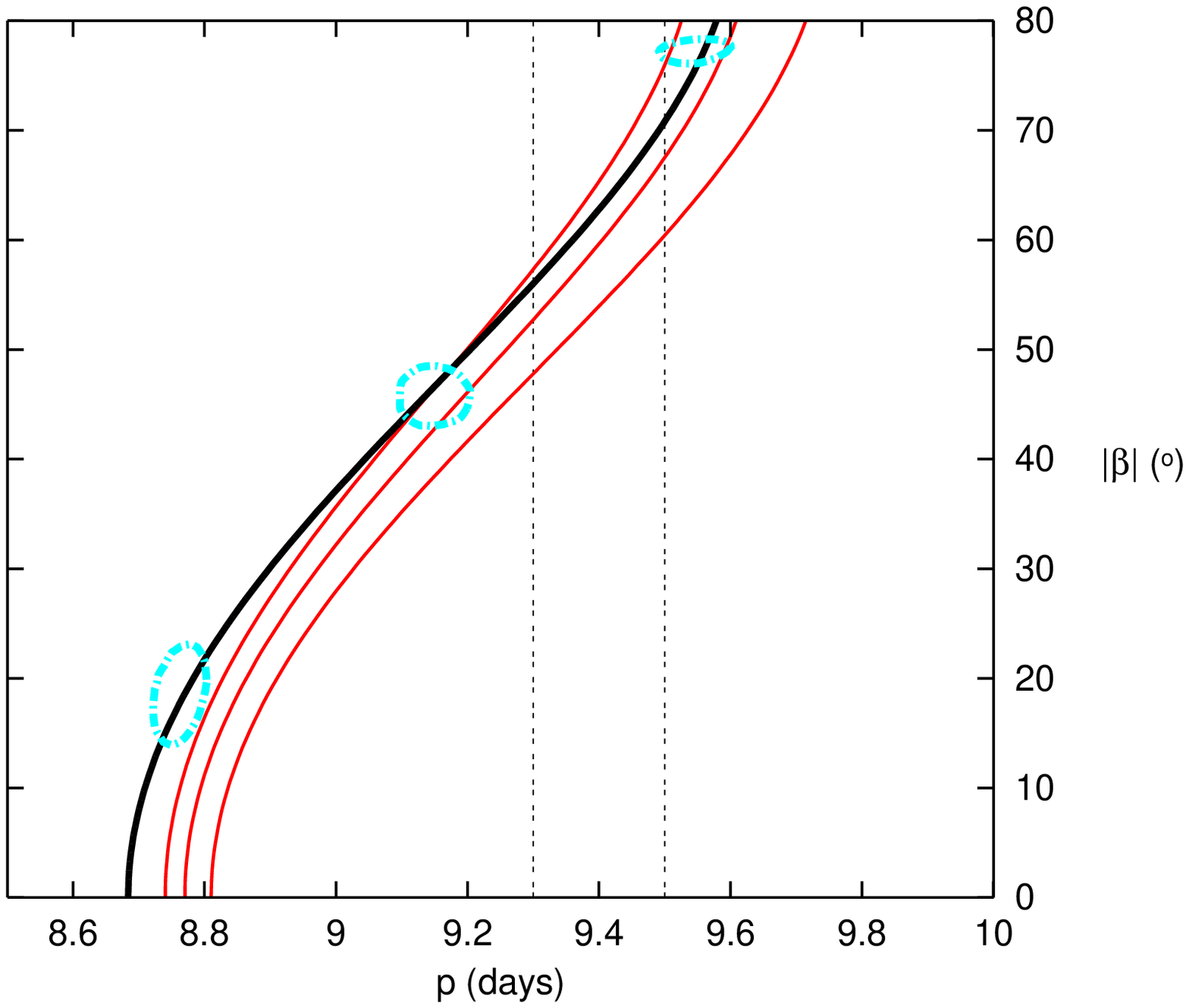}
%\plotone{contour_new_multiplot.eps}
\caption{
	{Top: The 68 \% marginalized likelihood contours for the modulus of the latitude, $|\beta|$, and spot period are shown by dots for each spot
	-- 2003 (green), 2004 (blue) and 2005 (pink). The central red curve is the solar-period, latitude relation (Equation 1) with the mean values
	taken from Solution 1 of Table 3, $k$ = \kBest \ and $P_{EQ}$ = \PEQBest$d$. 
	The other two red lines correspond to $k$ = \kLow, $P_{EQ}$ = \PEQLow$d$,
	and $k$ = \kHigh, $P_{EQ}$ = \PEQHigh$d$.
	The vertical dashed lines indicate the range of rotational periods ($p$ = 9.4 $\pm$ 0.1 days) found in Ca II H \& K line reversals over $\sim$30 years.
	Bottom: The 68 \% marginalized likelihood contours for the 2004 data only (cyan) while fitting not for $k$ and $P_{EQ}$ but explicitly for
	the periods, $p_{2004\_1}$, $p_{2004\_2}$, and $p_{2004\_3}$.
	The black curve is the best-fitting curve ($k$ = \kFourBest \ and $P_{EQ}$ = \PEQFourBest$d$)
	for the 2004 data, while the red curves are the same as in the upper panel.
	As can be seen, the black curve fits the data well and thus the observed differential rotation pattern
	is closely similar to solar.
	}
\label{FigPversusB}
}
\end{figure}

\end{document}